\documentclass[twocolumn,amsmath,amssymb,aps,pra]{revtex4-1}
\usepackage[english]{babel}
\usepackage[utf8]{inputenc}
\usepackage{graphicx}
\usepackage{mathtools}
\usepackage{bbm}

\usepackage{color}

\newcommand*{\mat}[1]{\boldsymbol{#1}}
\newcommand*{\coord}[1]{\mathbf{#1}}
\newcommand*{\vecr}{\coord{r}}

\newcommand*{\vecx}{\coord{x}}
\newcommand*{\vecy}{\coord{y}}

\newcommand*{\binteg}[3]{\int^{\mathrlap{#3}}_{\mathrlap{#2}}\ud{#1}\,}
\newcommand*{\integ}[1]{\int\!\!\!\:\ud{#1}\:}

\newcommand*{\crea}[1]{\hat{#1}^{\dagger}}
\newcommand*{\anni}[1]{\hat{#1}^{\vphantom{\dagger}}}

\newcommand{\ket}[1]{\lvert{#1}\rangle}

\newcommand{\brakket}[3]{\langle{#1}|{#2}|{#3}\rangle}
\newcommand{\bigbrakket}[3]{\bigl\langle{#1}\big|{#2}\big|{#3}\bigr\rangle}

\DeclareMathOperator{\area}{area}
\DeclareMathOperator{\Imag}{\mathfrak{Im}}
\DeclareMathOperator{\Laplace}{\mathcal{L}}

\newcommand*{\abs}[1]{\lvert#1\rvert}
\newcommand*{\du}{\partial}
\newcommand*{\e}{\textrm{e}}

\newcommand*{\half}{\frac{1}{2}}
\newcommand*{\I}{\mathrm{i}}
\newcommand*{\isDefinedAs}{\coloneqq}

\newcommand*{\Reals}{\mathbb{R}}

\newcommand*{\ud}{\mathrm{d}}
\newcommand*{\unitop}{\mathbbm{1}}

\usepackage{hyperref}

\begin{document}

\title{Invertibility of retarded response functions for Laplace transformable potentials: application to one-body reduced density matrix functional theory}
\author{K.J.H. Giesbertz}
\email{k.j.h.giesbertz@vu.nl}
\affiliation{Section of Theoretical Chemistry, Faculty of Exact Sciences, VU University, De Boelelaan 1083, 1081 HV Amsterdam, The Netherlands}

\begin{abstract}
A theorem for the invertibility of arbitrary response functions is presented under the following conditions: the time-dependence of the potentials should be Laplace transformable and the initial state should be a ground state, though it might be degenerate. This theorem provides a rigorous foundation for all density-functional-like theories in the time-dependent linear response regime. Especially for time-dependent one-body reduced density matrix (1RDM) functional theory this is an important step forward, since a solid foundation has currently been lacking. The theorem is equally valid for static response functions in the non-degenerate case, so can be used to characterize the uniqueness of the potential in the ground state version of the corresponding density-functional-like theory. Such a classification of the uniqueness of the non-local potential in ground state 1RDM functional theory has been lacking for decades. With the aid of presented invertibility theorem presented here, a complete classification of the non-uniqueness of the non-local potential in 1RDM functional theory can be given for the first time.
\end{abstract}

\maketitle

\section{Introduction}

The main statement of the Runge--Gross theorem is that the potential to density mapping is invertible modulo a time-dependent constant in the potential~\cite{RungeGross1984}. This statement implies that knowledge of the time-dependent density and the initial state is in principle sufficient to fully characterize the evolution of a quantum system. The invertibility theorem by Runge--Gross therefore forms the cornerstone of time-dependent density functional theory (TDDFT)~\cite{tddft2006, CasidaHuix-Rotllant2012, Ullrich2012, tddft2012}. Unfortunately, the Runge--Gross theorem only holds for potentials which are Taylor-expandable in time. 
Taylor expandability of the potential is actually a too stringent condition and can be loosened as demonstrated by the invertibility theorem for linear response by Van Leeuwen~\cite{Leeuwen2001} and more recently, by work of Tokatly on lattice systems~\cite{Tokatly2011a, Tokatly2011b, FarzanehpourTokatly2012} and the fixed-point approach by Ruggenthaler and Van Leeuwen~\cite{RuggenthalerLeeuwen2011, RuggenthalerGiesbertzPenz2012, RuggenthalerPenzLeeuwen2015}.

Practical TDDFT calculations are almost exclusively performed with the help of an auxiliary non-interacting reference system, the Kohn--Sham system~\cite{RungeGross1984, KohnSham1965}, which has the same density as the fully interacting system thanks to the exchange-correlation potential. The exact exchange-correlation potential depends on all densities at earlier times and on the initial states in a complicated manner~\cite{RuggenthalerNielsenLeeuwen2013}. This history dependence of the exact exchange-correlation potential is neglected in practice and replaced by its ground state version in which the instantaneous density is inserted. This neglect of memory dependence is known as the adiabatic approximation and the results are typically very satisfactory for polarizabilities and local valence excitations. Especially when a good model for the exchange-correlation potential is used even Rydberg excitations can be reproduced reliably~\cite{GisbergenKootstraSchipper1998, SchipperGritsenkoGisbergen2000, AppelGrossBurke2003}. Practical TDDFT calculations fail for more complicated excitations such as charge transfer excitations~\cite{DreuwWeismanHead-Gordon2003, GritsenkoBaerends2004b} and bound excitons~\cite{RohlfingLouie1998, BenedictShirleyBohn1998}, when the hole and electron are not localized close to each other~\cite{BaerendsGritsenkoMeer2013}, though some progress has been reported in their TDDFT description~\cite{ReiningOlevanoRubio2002, YanaiTewHandy2004, YangUllrich2013}. Even more problematic are double~\cite{MaitraZhangCave2004, CaveZhangMaitra2004} and bond-breaking~\cite{GritsenkoGisbergenGorling2000, GiesbertzBaerends2008} excitations. The main problem is that the density is not a natural quantity to describe these excitation processes and the non-interacting Kohn--Sham system is also of no avail.

A more natural quantity to deal with these more complicated physical processes is the one-body reduced density matrix (1RDM). In particular its fractional occupation numbers are good descriptors of correlation effects. Indeed, it has been demonstrated that time-dependent 1RDM functional theory is capable of correctly describing charge-transfer excitations, double excitations and bond-breaking excitations~\cite{GiesbertzBaerendsGritsenko2008, GiesbertzPernalGritsenko2009, GiesbertzGritsenkoBaerends2010a} even within the adiabatic approximation.
Unfortunately, no proper formal justification for time-dependent 1RDM functional theory has yet been published. The main purpose of this paper is to partially eliminate this caveat by presenting an invertibility theorem for non-local potentials and the 1RDM in the linear response regime.

In previous work~\cite{PernalGritsenkoBaerends2007} one tried to avoid the lack of a formal foundation by invoking the the Runge--Gross theorem. If all observables are functionals of the time-dependent density, they certainly are of the 1RDM, since the density can readily be extracted from the 1RDM. The problem is that there are (infinitely) many 1RDMs generating the same density, but only one of these 1RDMs corresponds to the local potential belonging to that density, the `local' 1RDM. So if we want to use the Runge--Gross theorem (or one of the extensions) as a foundation for time-dependent 1RDM functional theory, we are only allowed to use these `local' 1RDMs. Since the characterization of these `local' 1RDMs is seems to be impossible, direct use of the Runge--Gross theorem does not lead to any viable theory.

Actually, if we follow the philosophy of TDDFT more closely, we should not consider the mapping from local potentials to 1RDMs, but from non-local potentials to 1RDMs, since the non-local potential is the natural conjugate variable of the 1RDM. With a non-local potential I mean a one-body potential non-local in space, but still local in time. A generalization of the Runge--Gross theorem from local potentials and densities to non-local potentials and 1RDMs would therefore be more appropriate. Unfortunately, a straightforward generalization is not possible, since the commutator between the 1RDM and the interaction does not vanish, $\bigl[\hat{\gamma},\hat{W}\bigr] \neq 0$.

The invertibility theorem for the density response function for Laplace transformable potentials by Van Leeuwen~\cite{Leeuwen2001} is much more amenable to generalization to the 1RDM. Assuming that the initial state is a non-generate ground state and that the time-dependent part of the potential is Laplace transformable, the theorem by Van Leeuwen states that the density response function is invertible up to a constant shift in the potential. The proof by Van Leeuwen can be split into two parts. The first part consists of the derivation of a necessary and sufficient condition for a perturbation to yield zero response, i.e.\ necessary and sufficient condition for a potential to belong to the kernel of the response function. This first part does not need the special properties of the density operator and can therefore readily be generalized to arbitrary operators, e.g.\ the 1RDM operator, the current operator, the (non-collinear) spin density operator and the kinetic energy density operator. The second step of the proof by Van Leeuwen is to check the necessary and sufficient condition for the density operator, so the second part depends on the particular properties of the density. A generalization of the second step to arbitrary operators is therefore not possible, but needs to be considered separately for each operator. In this article I will work out the second step for the 1RDM operator, which allows me to completely characterize the kernel of the 1RDM response function.

The first part of the proof by Van Leeuwen requires at one point that the initial ground state is non-degenerate. This non-degeneracy requirement is quite a nuisance, since this requirement immediately excludes all open shell systems. I will not only repeat the first part of the proof by Van Leeuwen for arbitrary operators, but also show how the theorem can be extended to include degenerate initial states. Including initial degenerate ground states leads to an additional condition which needs to be checked. I work out this additional condition for the density operator and show that the inclusion of degenerate states does not lead to additional potentials in the kernel of the density response function. This provides an extension of the invertibility theorem for the density response function by Van Leeuwen to degenerate ground states. Degenerate ground states are also considered for the 1RDM response function.
Including the possibility of degeneracy leads to a more general expression for some of the potentials in the kernel of the 1RDM response function found non-degenerate case.
These results put linear response time-dependent 1RDM functional theory on a rigorous foundation. A more complete theoretical framework would be achieved by also addressing the $v$-representability question. The generalized invertibility theorem only determines to which extent the perturbations $\delta v$ yielding $\delta\gamma$ are unique. It does not answer the question if there actually exists a $\delta v$ which yields the desired $\delta\gamma$, i.e.\ the question of $v$-representability of $\delta\gamma$. For a complete theory, also the $v$-representability question should be addressed to specify exactly which $\delta\gamma$ we are allowed to use in linear response time-dependent 1RDM functional theory. This paper focusses on the invertibility of the response function, so the $v$-representability question is beyond the scope of this article.

As Van Leeuwen showed in Ref.~\cite{Leeuwen2003}, his invertibility proof for the density response function also works in the time-independent case by taking the limit of the variable in the Laplace transform to zero. Especially for ground state 1RDM functional theory this is an important result, since a Hohenberg--Kohn like proof for the non-local potential to 1RDM mapping does not exist as pointed out by Gilbert~\cite{Gilbert1975}. Using the invertibility proof for the 1RDM response function for the time-independent case I can give a full classification of the non-uniqueness of the non-local potential featured in 1RDM functional theory for the first time. Application to the time-independent case only works for non-degenerate ground states unfortunately, since degeneracies are treated in a fundamentally different manner in both cases.

The paper is organized as follows. I start by repeating the first part of the invertibility proof by Van Leeuwen for arbitrary operators and extend his approach to handle degenerate ground states as well. After necessary and sufficient conditions have been obtained to characterize the potentials which do not lead to a response, I work these conditions out for the density and the 1RDM operators. First I consider the density operator to check against the result by Van Leeuwen~\cite{Leeuwen2003} and to extend his result for the density response function to degenerate ground states. Next, I consider the 1RDM operator to find the non-local potentials which do not lead to a response of the 1RDM. First I consider the simpler case of non-degenerate ground states and then take the additional necessary condition into account to extend the result to degenerate ground states. In the last part before concluding, I discuss the implications of these results for ground state 1RDM functional theory and point out in more detail why degeneracies need a different treatment in the ground state case.

\section{The generalized invertibility theorem}
Now let us repeat the first part of the invertibility proof by Van Leeuwen~\cite{Leeuwen2001} for arbitrary operators $\hat{Q}_i$ and generalize the proof to degenerate ground states. The set of operators $\{\hat{Q}_i\}$ can be any set of self-adjoint operators of interest, e.g.\ dipole and quadrupole operators. The index $i$ is also allowed to be a continuous index to represent a self-adjoint operator density such as the density operator, $\hat{n}(\vecr)$. Of course, a mixture of continuous and discrete indices is also allowed such as in the spin-density operator, $\hat{n}(\vecx)$, or the 1RDM operator $\hat{\gamma}(\vecx,\vecx')$, where $\vecx \isDefinedAs \vecr\sigma$ is a combined space and spin coordinate. I will use the symbol $\isDefinedAs$ throughout the paper to emphasize definitions.

Having selected some set of self-adjoint operators and\slash{}or operator densities of interest, we consider perturbations by these operator $\hat{Q}_j$ with strengths $\delta v_j(t')$. Note that we need $\delta v_j(t') \in \Reals$ to ensure that the total perturbation remains hermitian. Now we consider the linear response of the expectation values of the same set of operators~\cite{FetterWalecka1971}
\begin{align}\label{eq:lineResp}
\delta Q_i(t) = \sum_j\binteg{t'}{0}{t} \chi_{ij}(t-t')\delta v_j(t') ,
\end{align}
where $\chi_{ij}(t-t')$ is the retarded\slash{}causal linear response function which can be defined as
\begin{align}\label{eq:respFuncDef}
\chi_{ij}(t-t') \isDefinedAs -\I\theta(t - t') \brakket{\Psi_0}{[\hat{Q}_{H_0,i}(t),\hat{Q}_{H_0,j}(t')]}{\Psi_0} .
\end{align}
In its definition we have used the operators in their Heisenberg representation with respect to the unperturbed Hamiltonian, $\hat{Q}_{H_0,i}(t) \isDefinedAs \e^{\I\hat{H}_0t}\hat{Q}_i\e^{-\I\hat{H}_0t}$, which is often referred to as the interaction picture. We have also used the Heaviside function, which is defined as
\begin{align*}
\theta(x) \isDefinedAs \begin{cases}
1 &\text{for $x > 0$} \\
0 &\text{for $x < 0$}.
\end{cases}
\end{align*}
The retarded response function can alternatively be expressed as a sum-over-states (its Lehmann representation) as~\cite{Lehmann1954, FetterWalecka1971, Leeuwen2003}
\begin{align}\label{eq:sosRespFunc}
\chi_{ij}(t-t') 
&= \I\theta(t-t')\sum_K\e^{\I\Omega_K(t-t')}{q_i^K}^*\!q_j^K + \text{c.c.} ,
\end{align}
where $\Omega_K \isDefinedAs E_K - E_0 \geq 0$ are excitation energies and $\Omega_K = 0$ only for $K < D$, so $D$ denotes the multiplicity of the ground state degeneracy~\footnote{The sum runs over a complete set of states, so also includes a possible continuum where the sum should be interpreted as an integral.}. Further, we have defined
$q^K_i \isDefinedAs \brakket{\Psi_0}{\hat{Q}_i}{\Psi_K}$.
Note that the initial state can be excluded from the sum, since $q_i^0 = \brakket{\Psi_0}{\hat{Q}_i}{\Psi_0} \in \Reals$, because the operators $\hat{Q}_i$ should be hermitian. Inserting the sum-over-state expression for the response function in~\eqref{eq:lineResp}, the response of the expectation value of the operator $\hat{Q}_i(t)$ can now be written as
\begin{align*}
\delta Q_i(t) = \I\sum_K{q^K_i}^*\binteg{t'}{0}{t} a_K(t')\e^{\I\Omega_K(t-t')} + \text{c.c.} ,
\end{align*}
where we have defined
\begin{align}\label{eq:aKdef}
a_K(t) \isDefinedAs \sum_jq^K_j\delta v_j(t) .
\end{align}
The integral has the form of a convolution product over the interval $[0,t]$ and can be transformed into a normal product by taking the Laplace transform
\begin{align*}
\Laplace [\delta Q_i](s) = \I\sum_K{q^K_i}^*\frac{\Laplace[a_K](s)}{s - \I\Omega_K} + \text{c.c.} ,
\end{align*}
where the Laplace transform is defined as
\begin{align*}
\Laplace [f](s) \isDefinedAs \binteg{t}{0}{\infty} \e^{-st}f(t) .
\end{align*}
Now we multiply this equation by the Laplace transform of the potential $\Laplace[\delta v_i](s)$ and sum over the index $i$ to obtain
\begin{align*}
\sum_i\Laplace [\delta v_i](s)\Laplace [\delta Q_i](s)
= -2\sum_K\frac{\Omega_K}{s^2 + \Omega_K^2}\abs{\Laplace [a_K](s)}^2 .
\end{align*}
In absence of response, we have that $\delta Q_i = 0$, so we also have that $\Laplace [\delta Q_i] = 0$ and we obtain from the previous equation that for zero response we necessarily have
\begin{align*}
0 = \sum_K\frac{\Omega_K}{s^2 + \Omega_K^2}\abs{\Laplace [a_K](s)}^2 .
\end{align*}
Because $\Omega_K \geq 0$ and only for $K < D$ we have $\Omega_K = 0$, all the contributions for $K \geq D$ are positive. Therefore, one necessarily has $\Laplace [a_K](s) = 0$ for $K \geq D$, so $a_K(t) = 0$ for $K \geq D$ as well. Strictly speaking, we should say that $a_K(t) = 0$ almost everywhere for $K \geq D$, since $a_K(t) \neq 0$ on a set of measure zero in time would not contribute to the integral of the Laplace transform.
From its definition~\eqref{eq:aKdef} it is clear that there are only two possibilities for $a_K(t) = 0$ almost everywhere. The first possibility is the absence of a perturbation, $\delta v_j(t) = 0$ almost everywhere. If we further assume that we are only interested in the classical solutions of the Schrödinger equation (these are the usual physical wave functions defined at each point in time), we have as an additional condition that the potential needs to be continuous up to its first order derivative in time, $\delta v_j(t) \in C^1$~\footnote{In Ref.~\cite{RuggenthalerPenzLeeuwen2015} it is stated that the condition $\delta v_j(t) \in C^1$ can probably be weakened to Lipschitz continuity. This is still sufficient for our argument, since we only need continuity. A milder version of the Schrödinger equation would allow for more general potentials in some $L^p$ spaces in time~\cite{PenzRuggenthaler2015, RuggenthalerPenzLeeuwen2015}. In that case, however, potentials which only differ at a set of zero measure would be considered equivalent.}, so `almost everywhere’ could be dropped. For more details on the solvability of the time-dependent Schrödinger equation, I refer the reader to an excellent introduction in Ref.~\cite{RuggenthalerPenzLeeuwen2015}. This first possibility is trivial, and will be excluded from further discussion.
The other possibility is that there exists one or more linear combinations of the operators under consideration
\begin{align*}
\hat{L}_n = \sum_j\hat{Q}_j\delta v_j^n,
\end{align*}
such that
$\brakket{\Psi_0}{\hat{L}_n}{\Psi_K} = 0$ for all $K \geq D$. Note that the linear combination should remain hermitian, so $\delta v_j^n \in \Reals$.
This implies that such a linear combination acting on the initial state, $\hat{L}_n\ket{\Psi_0}$, should not produce any components outside the degenerate subspace, i.e.
\begin{align}\label{eq:f0requirement}
\hat{L}_n\ket{\Psi_0} = \sum_{K < D}{l_n^K}^*\ket{\Psi_K} ,
\end{align}
where $l^K_n \isDefinedAs \brakket{\Psi_0}{\hat{L}_n}{\Psi_K}$.
In the case of a non-degenerate ground state the situation simplifies to an eigenvalue condition
\begin{align}\label{eq:eigenCond}
\hat{L}_n\ket{\Psi_0} = l_n\ket{\Psi_0} .
\end{align}
In words, for a non-degenerate initial ground state, the response can only be zero nontrivially, if there exists a linear combination of the operators $\hat{Q}_j$ for which the initial state is an eigenstate. Note $\ket{\Psi_0}$ being an eigenstate of $\hat{L}_n$ is sufficient, though not necessary for degenerate ground states, since $\hat{L}_n\ket{\Psi_0}$ is still allowed to have components in the degenerate subspace~\eqref{eq:f0requirement}.

Though we have shown that $a_K(t) = 0$ for $K \geq D$ is necessary for absence of response, we also need to check if this condition is sufficient.
Now suppose that condition~\eqref{eq:f0requirement} holds for some initial state $\ket{\Psi_0}$ and some operator $\hat{Q}_n$.
Note that we can always make a linear transformation of the set of operators such that the operators $\hat{L}_n$ are all explicitly contained in the set $\{\hat{Q}_n\}$.
In that case only the states with $\Omega_K = 0$ will contribute to the sum-over-state expression in~\eqref{eq:sosRespFunc}, so reduces to
\begin{align*}
\chi_{ni}(t-t') &=
\I\theta(t-t')\sum_K \Imag\bigl[{q_n^K}^*q_i^K\bigr] +\text{c.c.} \\
&= \I\theta(t-t')\brakket{\Psi_0}{[\hat{Q}_i,\hat{Q}_n]}{\Psi_0} .
\end{align*}
so as an additional requirement for zero response apart from $q^K_n = 0$ for $K \geq D$, we find that
\begin{align}\label{eq:fDegenComm}
\brakket{\Psi_0}{[\hat{Q}_i,\hat{Q}_n]}{\Psi_0} = 0 \qquad \forall_i .
\end{align}
A number of remarks on this condition are in order. If the initial state is an eigenstate of the operator $\hat{Q}_n$, this condition is automatically satisfied. Hence, it is sufficient for non-degenerate ground states to check condition~\eqref{eq:eigenCond}.
This implies that only in the case of a degenerate initial state for which $\hat{Q}_n\ket{\Psi_0}$ has some components in the degenerate subspace, condition~\eqref{eq:fDegenComm} needs to considered explicitly. Since operators commute with themselves, condition~\eqref{eq:fDegenComm} is trivially satisfied for $i=n$, so the check only needs to be performed for $i \neq n$.

\textit{Example.}
To get a feeling how condition~\eqref{eq:fDegenComm} comes into play, consider a three dimensional Hilbert space, $\mathcal{H} = \{\ket{2s},\ket{2p_x},\ket{2p_y}\}$.
We take the Hamiltonian of the hydrogen atom as our initial Hamiltonian, $\hat{H}_0$, so the states in $\mathcal{H}$ are degenerate.
The condition~\eqref{eq:f0requirement} is therefore trivially satisfied and only the additional condition~\eqref{eq:fDegenComm} needs to be considered.
For the operators $\hat{Q}_j : \mathcal{H} \to \mathcal{H}$ we consider the unit operator and the dipole operators in the $x$- and $y$-direction, $\{\hat{Q}_j\} = \{\unitop, x, y\}$.
We select the $2s$-orbital to be our initial state, $\ket{\Psi_0} = \ket{2s}$.
Since any state is an eigenstate of the unit operator, we immediately find that at least the unit operator is a part of the kernel of the response function. The dipole operators produce an additional component in the degenerate subspace~\footnote{The matrix elements {$\ket{\psi_i}\brakket{\psi_i}{\vecr}{2s}$} have been evaluated by calculating the corresponding integral.}
\begin{align*}
x\ket{2s} &\isDefinedAs \sum_{\crampedclap{\psi_i \in \mathcal{H}}}\ket{\psi_i}\brakket{\psi_i}{x}{2s} = -3\ket{2p_x} ,	\\
y\ket{2s} &\isDefinedAs \sum_{\crampedclap{\psi_i \in \mathcal{H}}}\ket{\psi_i}\brakket{\psi_i}{y}{2s} = -3\ket{2p_y} ,
\end{align*}
so the additional condition~\eqref{eq:fDegenComm} needs to be checked explicitly. The dipole operators already commute among themselves, $[x,y] = 0$, so also the dipole operators belong to the kernel of the response function. Because all operators belong to the kernel of the response function, we actually have $\mat{\chi} = \mat{0}$. Such a response function is rarely encountered in practice and is a consequence of the special choice of the Hilbert space, $\mathcal{H}$, and the set of operators $\{\unitop, x, y\}$. If, for example, also the $\ket{3p_x}$ state would be included in the Hilbert space, the $x$-operator would produce also components outside the degenerate subspace
\begin{align*}
x\ket{2s} = \sum_{\crampedclap{\psi_i \in \mathcal{H} \cup \{\ket{3p_x}\}}}\ket{\psi_i}\brakket{\psi_i}{x}{2s}
=  \frac{27\,648}{15\,625}\ket{3p_x} - 3\ket{2p_x} .
\end{align*}
By the first condition~\eqref{eq:f0requirement}, the $x$-operator would not belong to the kernel of the response function anymore. Note that the extension has only consequences for the $x$-operator, so the unit-operator and $y$-operator remain in the kernel of the response function.

The $x$-operator can also be lifted out of the kernel of the response function by adding additional operators. For example, consider an extension of the set of operators by an operator which applies the momentum operator three times in the $x$-direction, $\I\du_x^3 = (-\I\du_x)^3$~\footnote{The momentum operator gives {$-\I\du_x\ket{2s} = 0$} in the limited Hilbert space $\mathcal{H}$, so is simply the zero-operator. Using $(-\I\du_x)^3$ avoids such a pathological operator.}. The action of $\I\du_x^3$ on the initial state is defined to be
\begin{align*}
\I\du_x^3\ket{2s} \isDefinedAs \sum_{\crampedclap{\psi_i \in \mathcal{H}}}\ket{\psi_i}\brakket{\psi_i}{\I\du_x^3}{2s}
= \frac{\I}{20}\ket{2p_x} .
\end{align*}
Consider now the additional condition~\eqref{eq:fDegenComm} for $\I\du_x^3$. Only the dipole operator in the $x$-direction yields a non-vanishing commutator and its expectation value for the initial state gives
\begin{align*}
\bigbrakket{2s}{\bigl[x,\I\du_x^3\bigr]}{2s} = \frac{\I}{4} \neq 0 .
\end{align*}
The operator $\I\du_x^3$ is therefore not part of the kernel of the `extended' response function. Note that this result also implies that for the dipole operator in the $x$-direction condition~\eqref{eq:fDegenComm} is not satisfied anymore. Only the unit operator and the dipole operator in the $y$-direction span therefore the kernel of the `extended' response function.

\section{Density response}
As a minor check, let us consider the density response function to see if we recover the original result by Van Leeuwen~\cite{Leeuwen2001}. For the density response function our operators are $\hat{Q}_{\vecr} = \hat{n}(\vecr)$, where $\hat{n}(\vecr) \isDefinedAs \sum_{\sigma}\crea{\psi}(\vecr\sigma)\anni{\psi}(\vecr\sigma)$. The only linear combination for which a non-degenerate ground state is an eigenstate is the number operator
\begin{align}\label{eq:totalNumberOperators}
\hat{N} \isDefinedAs \integ{\vecr}\hat{n}(\vecr) .
\end{align}
Only if the density would vanish in some region for $\ket{\Psi_0}$, there would be other linear combinations for which $\ket{\Psi_0}$ would be an eigenstate. This possibility is typically excluded in DFT~\cite{HohenbergKohn1964, ReedSimon1980, Lieb1983, Lammert2015}, so we recover the same result as Van Leeuwen~\cite{Leeuwen2001} that only a spatially constant potential gives a zero density response.

Now let us investigate the consequences of a degenerate ground state by considering only one particle first. A non-constant potential yielding a zero density response is readily constructed as 
$v_K(\vecr) = \Psi_K(\vecr) / \Psi_0(\vecr)$ for $0 < K < D$,
which by construction satisfies~\eqref{eq:f0requirement}. This construction only works if the wave function $\Psi_K(\vecr)$ has no imaginary part. Otherwise the potential $v_K$ would not be hermitian.
Assuming that the ground states $\Psi_K(\vecr)$ are indeed real, the initial state should at least have one nodal surface to allow for the degeneracy. Further, because the states need to be orthogonal, not all of their nodal surfaces should coincide. The potential $v_K$ would therefore be infinite along some nodal surface of $\Psi_0$~\footnote{An example is the hydrogen atom where the $1s$ state is excluded from the Hilbert space. Choose the $2p_x$ orbital as an initial state. The local potential which would produce only a component in the $2p_y$ state would be $y/x$ which is infinite in the whole plane orthogonal to the $x$-axis.}.

For the next step we need to take additional conditions on the allowed perturbation into account, to ensure that the Hamiltonian remains self-adjoint~\cite{ReedSimon1975, RuggenthalerPenzLeeuwen2015}. For potentials over $\Reals^3$ one requires the perturbations to be in the class of Kato potentials $\mathcal{K} \isDefinedAs L^2(\Reals^3) + L^{\infty}(\Reals^3)$~\cite{Kato1957}. This means that the perturbing potential needs to be decomposable in parts which are either bound or square (Lebesgue) integrable. For example the Coulomb potential can be split into two parts as
\begin{align*}
\frac{1}{\abs{\vecr}} = \frac{\theta(1 - \abs{\vecr})}{\abs{\vecr}} + \frac{\theta(\abs{\vecr} - 1)}{\abs{\vecr}} .
\end{align*}
The first part contains the Coulomb singularity, which that is square-integrable, so the first part belongs to $L^2$. The second part is the outer region of the Coulomb potential which is bounded (between 0 and 1), so belongs to $L^{\infty}$. The Coulomb potential is therefore a proper perturbative potential, since it is a Kato perturbation. An example of a potential which is not Kato is the harmonic oscillator potential, $\half\omega^2\abs{\vecr}^2$. Due to its divergence for $\abs{\vecr} \to \infty$ the outer part of the harmonic potential is neither bound nor square integrable.
Returning to our potentials $v_K(\vecr) = \Psi_K(\vecr) / \Psi_0(\vecr)$, we expect these potentials to behave as $1/z$ in the direction orthogonal to the nodal surface. These potentials are therefore expected not to be square integrable in regions including some part of the nodal surface. To proof this suspicion, we use the result by Kato~\cite{Kato1957} that the solutions of the time-independent Schrödinger equation are continuous over whole $\Reals^{3N}$ and their derivatives locally in $L^{\infty}$, i.e.\ the solutions are locally Lipschitz. Locally Lipschitz means that for each $\vecx_0,\vecy_0 \in \Reals^{3N}$ there exists some neighborhood $V$ and a constant $K_V$ such that
$\abs{\Psi(\vecx) - \Psi(\vecy)} \leq K_V\abs{\vecx - \vecy}$ for any $\vecx,\vecy \in V$. Now take a neighborhood around some point at the nodal surface of $\Psi_0(\vecr)$ and call this neighborhood $S$. The local Lipschitz condition for $\Psi_0$ simplifies in this neighborhood to $\abs{\Psi_0(s)} \leq K_S\abs{s}$ where $s$ denotes the distance from the nodal surface. The sign of $s$ indicates on which side of the surface we are. Since the nodal surfaces of $\Psi_K(\vecr)$ and $\Psi_0(\vecr)$ do not coincide, we can always choose some neighborhood such that $\min(\abs{\Psi_K}) = E > 0$. The $L^2$-norm of the potential within $S$ can now be estimated as
\begin{align*}
\binteg{\vecr}{S}{}\biggl\lvert\frac{\Psi_K(\vecr)}{\Psi_0(\vecr)}\biggr\rvert^{\mathrlap{2}}
&\geq \biggl(\frac{E}{K_S}\biggr)^{\mathclap{2}}\binteg{\vecr}{S}{}\frac{1}{s^2} \\
&\geq \biggl(\frac{E}{K_S}\biggr)^{\mathclap{2}}\area(S^*)\binteg{s}{-\epsilon}{\epsilon}\frac{1}{s^2} = \infty ,
\end{align*}
where $\area(S^*) > 0$ and $\epsilon > 0$.
In the last inequality, $S^*$ is a part of the nodal surface contained in $S$, such that $S^* \times [-\epsilon,\epsilon] \subseteq S$. Thus the last integral is basically over a slab of thickness $2\epsilon$ contained in $S$ along the nodal surface. The inequality therefore shows that the potential $v_K(\vecr)$ is not square integrable in the region $S$, so this part of the potential is not in $L^2$. Since the potential is obviously not bounded in this region, i.e.\ not in $L^{\infty}$, the potential $v_K(\vecr)$ is not a Kato perturbations. The potentials $v_K(\vecr)$ are therefore not admissible, so degenerate initial states do not form any complication for the invertibility of the density response function. The final result is that also in the degenerate case the kernel of the density response function only consists of a constant potential. It is obvious that the same conclusion also holds for more than one particle.

\section{1RDM response}
We will now consider the 1RDM response function. The 1RDM operator is defined as
\begin{align*}
\hat{\gamma}(\vecx,\vecx') \isDefinedAs \crea{\psi}(\vecx')\anni{\psi}(\vecx) ,
\end{align*}
where $\anni{\psi}(\vecx)$ and $\crea{\psi}(\vecx)$ are the usual field operators and $\vecx \isDefinedAs \vecr\sigma$ is a combined space-spin coordinate.
We will first limit ourselves to a non-degenerate ground state as initial state, since this case already leads to several situations which need to be considered.

\subsection{Non-degenerate ground state as initial state}
\label{sec:nondegenerate1RDM}

Because the density is simply the diagonal of the 1RDM, $n(\vecr) = \sum_{\sigma}\gamma(\vecr\sigma,\vecr\sigma)$, the constant potential is also present in the kernel of the 1RDM response function.
Since the 1RDM contains more flexibility than the density, one would expect that there are more possible potentials that give a zero response than only the spatially constant potential. Indeed, any one-body operator can be represented by the 1RDM, so if the initial state is an eigenfunction of some one-body operator, this operator is also present in the kernel of the 1RDM response function.

In particular, the non-relativistic Hamiltonian does not depend on spin, so a non-degenerate ground state is necessarily a singlet state. This implies that the ground state is an eigenstate of the total spin-projection operator in arbitrary directions, $\hat{\mat{S}} \ket{\Psi_0} = 0$. Since the total spin-projection operator can be expressed as a one-body operator, it is also part of the kernel of the 1RDM response function. Note that this situation also occurs in spin-DFT~\cite{BarthHedin1972, EschrigPickett2001}.

Since symmetry in the system implies that the Hamiltonian commutes with one or more symmetry operators, the eigenstates of the Hamiltonian can be chosen to be eigenstates of some of those symmetry operators as well. Therefore, one would expect that also these symmetry operators belong to the kernel of the 1RDM response function. However, the Coulomb interaction of the Hamiltonian couples all the particles, so these symmetry operators need to be many-body operators in general. Continuous symmetries form an exception, since their generators can be expressed as one-body operators. For linear molecules this would be the rotation around the $z$-axis, i.e.\ the $\hat{L}_z$ operator. Atoms would also include the other total angular momentum operators, $\hat{L}_x$ and $\hat{L}_y$. For systems which are homogeneous in one or more directions, e.g.\ the homogeneous electron gas, the corresponding momentum operator(s) would also be part of the kernel of the 1RDM response function.

To proceed with the analysis, we will work in the natural orbital (NO) basis of the 1RDM of the initial ground state, which can be obtained by diagonalizing the 1RDM
\begin{align*}
\gamma(\vecx,\vecx') = \sum_kn_k\,\phi_k(\vecx)\phi_k^*(\vecx') .
\end{align*}
The eigenvalues are called the (natural) occupation numbers and the eigenfunctions are the natural orbitals~\cite{Lowdin1955}. The occupation numbers sum to the total number of particles in the system, $N$, and for fermions they obey $0 \leq n_k \leq 1$. The integer values are special, since $n_k = 0$ implies that the NO $\phi_k(\vecx)$ is not present in any determinant in the expansion of the wavefunction, $\anni{a}_k\ket{\Psi_0} = 0$, where $\anni{a}_k$ is the annihilation operator for the NO $\phi_k(\vecx)$.
Likewise, a fully occupied NO, $n_k = 1$, implies that the NO $\phi_k(\vecx)$ is present in all determinants, so $\crea{a}_k\anni{a}_k\ket{\Psi_0} = \ket{\Psi_0}$~\cite{Lowdin1955}. From these properties, we readily find that
\begin{align*}
\hat{\gamma}_{k,l}\ket{\Psi_0} = \begin{cases}
0\ket{\Psi}							&\text{if $n_k = 0$ or} \\
								&\text{\hphantom{if }$n_l = 1$ and $k \neq l$} \\
1\ket{\Psi}							&\text{if $n_l = 1$ and $k = l$} \\
{\displaystyle \sum_K}c_K\ket{\Psi_K}	&\text{otherwise} ,
\end{cases}
\end{align*}
where the 1RDM operator is now represented in the NO basis, $\hat{\gamma}_{k,l} \isDefinedAs \crea{a}_l\anni{a}_k$.
Hence we find that the ground state is an eigenstate of the 1RDM operator if $n_k = 0$ or $n_l = 1$. However, we have to keep in mind that the potential should be hermitian, so if $\delta v_{kl} \neq 0$, also $\delta v^*_{lk} \neq 0$. Thus for the state $\ket{\Psi}$ to be an eigenstate of both $\hat{\gamma}_{k,l}$ and $\hat{\gamma}_{l,k}$, we additionally need that $n_k = 1$ or $n_l = 0$. This situation can only occur if $n_k = n_l = 0$ or $n_k = n_l = 1$. We find therefore, that the perturbations within the fully occupied block or within the completely unoccupied block have a zero response in the 1RDM, as is actually well known for non-interacting systems, e.g.\ the Kohn--Sham system in DFT.
Note that this discussion includes the one-particle case, since that is also non-interacting.

For interacting systems, the occupation numbers are predominantly fractional, $0 < n_k < 1$, and for Coulomb systems there is strong evidence that they all are~\cite{Friesecke2003, GiesbertzLeeuwen2013a, GiesbertzLeeuwen2013b, GiesbertzLeeuwen2014}. One would expect that another special situation can occur if these fractional occupation numbers are degenerate. To investigate this situation, consider the NOs as a basis and assume that $\phi_1(\vecx)$ and $\phi_2(\vecx)$ are two degenerate NOs. The contribution of these degenerate NOs to the initial state can be made explicit by writing the initial state as
\begin{align*}
\ket{\Psi_0} = \crea{a}_1\crea{a}_2\ket{\widetilde{\Psi}^{12}_{N-2}} + 
\crea{a}_1\ket{\widetilde{\Psi}^1_{N-1}} +
\crea{a}_2\ket{\widetilde{\Psi}^2_{N-1}} + \ket{\widetilde{\Psi}_N} ,
\end{align*}
where $\anni{a}_i\ket{\widetilde{\Psi}^b_M} = 0$ for $i = 1,2$ and any $b \in \{\emptyset, 1, 2, 12\}$.
Note that the states $\ket{\widetilde{\Psi}^a_M}$ are not normalized in general.
The action of the 1RDM-operator on the initial state can be worked out as
\begin{align*}
\hat{\gamma}_{i,j}\ket{\Psi_0}
= \crea{a}_j\ket{\widetilde{\Psi}^i_{N-1}} + \delta_{i,j}\crea{a}_1\crea{a}_2\ket{\widetilde{\Psi}^{12}_{N-2}}
\end{align*}
for $i,j = 1,2$. Since the $\ket{\widetilde{\Psi}_N}$ component vanishes, the only way that $\ket{\Psi_0}$ can be an eigenstate is to have the eigenvalue zero, so all components $\ket{\widetilde{\Psi}^b_M}$ need to be cancelled. The components $\crea{a}_j\ket{\widetilde{\Psi}^i_{N-1}}$ are not present in the initial state $\ket{\Psi_0}$. This follows from the fact that $\phi_1(\vecx)$ and $\phi_2(\vecx)$ are NOs, so $\gamma_{12} = 0$. Since one generally can not eliminate these components by taking linear combinations of $\hat{\gamma}_{i,j}$, fractional occupation degeneracies do \emph{not} cause additional potentials in the kernel of the 1RDM response function in general.

A special situation occurs if $\ket{\widetilde{\Psi}^i_{N-1}} = 0$. The only known interacting case is the two-electron system. The two-electron state in the NO representation can be written as an expansion of NO pairs to which each NO contributes only once~\cite{LowdinShull1956, CioslowskiPernalZiesche2002, PhD-Giesbertz2010, RappBricsBauer2014}
\begin{align*}
\ket{\Psi_0} = \sum_{k=1}^{\infty}c_k\,\crea{a}_{\vphantom{\bar{k}}k}\crea{a}_{\bar{k}}\ket{} .
\end{align*}
The coefficients in the expansion are called natural amplitudes and are related to the occupation numbers as $\abs{c_k}^2 = n_k = n_{\bar{k}}$. In the case of a singlet state, the NO pairs only differ in their spin part, $\phi_k(\vecx) = \phi_k(\vecr)\alpha(\sigma)$ and $\phi_{\bar{k}}(\vecx) = \phi_k(\vecr)\beta(\sigma)$. In the triplet case the NO pairs have different spatial parts and their spin parts can be identical. The paired NOs are degenerate and since we now have $\ket{\widetilde{\Psi}^k_{N-1}} = 0$, we find that
\begin{align}\label{eq:pairedZeroResponse}
0 = \hat{\gamma}_{k,\bar{k}}\ket{\Psi_0} = \hat{\gamma}_{\bar{k},k}\ket{\Psi_0}
= \bigl(\hat{\gamma}_{k,k} - \hat{\gamma}_{\bar{k},\bar{k}}\bigr)\ket{\Psi_0} ,
\end{align}
so perturbations with these operators yield a zero 1RDM response~\footnote{If $\Psi_0$ is a singlet state,~\eqref{eq:pairedZeroResponse} are triplet operators.}.

The special structure of the two-electron state also causes other NOs with degenerate natural occupation numbers to yield zero response. For example, consider the contribution of two pairs of NOs to the initial state
\begin{align*}
c_1\crea{a}_{\vphantom{\bar{1}}1}\crea{a}_{\bar{1}}\ket{} + c_2\crea{a}_{\vphantom{\bar{2}}2}\crea{a}_{\bar{2}}\ket{} .
\end{align*}
Now we work out the action of the following perturbations which mix these NO pairs
\begin{align*}
\bigl(v_{21}\hat{\gamma}_{1,2} + v_{21}^*\hat{\gamma}_{2,1}\bigr)\ket{\Psi_0}
&= v_{21}c_1\crea{a}_{\vphantom{\bar{1}}2}\crea{a}_{\bar{1}}\ket{} +
v_{21}^*c_2\crea{a}_{\vphantom{\bar{2}}1}\crea{a}_{\bar{2}}\ket{} , \\
\bigl(v_{\bar{1}\bar{2}}\hat{\gamma}_{\bar{2},\bar{1}} + v_{\bar{1}\bar{2}}^*\hat{\gamma}_{\bar{1},\bar{2}}\bigr)\ket{\Psi_0}
&= v_{\bar{1}\bar{2}}\,c_2\crea{a}_{\vphantom{\bar{1}}2}\crea{a}_{\bar{1}}\ket{} + 
v_{\bar{1}\bar{2}}^*\,c_1\crea{a}_{\vphantom{\bar{2}}1}\crea{a}_{\bar{2}}\ket{} .
\end{align*}
Note that we added a second term to each perturbation, to ensure that the operator is hermitian.
We see that both perturbations produce exactly the same determinants, so by combining both perturbations we might be able to cancel both. To eliminate the first determinant, $\crea{a}_2\crea{a}_{\bar{1}}\ket{}$, we need to set $v_{\bar{1}\bar{2}} = -v_{21}c_1/c_2$. To eliminate the second determinant, $\crea{a}_1\crea{a}_{\bar{2}}\ket{}$, we need to set $v^*_{\bar{1}\bar{2}} = -v^*_{21}c_2/c_1$. This only works when the natural occupations are degenerate, $n_1 = \abs{c_1}^2 = \abs{c_2}^2 = n_2$. In that case the following potential belongs to the kernel of the 1RDM response function
\begin{align}\label{eq:kernPotDegen1}
v_{21}\bigl(\hat{\gamma}_{1,2} - \e^{\I\alpha_{12}} \gamma_{\bar{2},\bar{1}}\bigr) +
v_{21}^*\bigl(\hat{\gamma}_{2,1} - \e^{-\I\alpha_{12}} \gamma_{\bar{1},\bar{2}}\bigr) ,
\end{align}
which depends on the relative phase of the natural amplitudes, $\e^{\I\alpha_{12}} \isDefinedAs c_1/c_2$.
It is readily checked that degeneracy between the NO pairs implies that also the potential
\begin{align}\label{eq:kernPotDegen2}
v_{\bar{2}1}\bigl(\hat{\gamma}_{1\bar{2}} + \e^{\I\alpha_{12}} \hat{\gamma}_{2\bar{1}}\bigr) +
v_{\bar{2}1}^*\bigl(\hat{\gamma}_{\bar{2}1} + \e^{-\I\alpha_{12}} \hat{\gamma}_{\bar{1}2}\bigr)
\end{align}
belongs to the kernel of the response function. Note that the relative phase of the natural amplitude is important in the construction of the potentials~\eqref{eq:kernPotDegen1} and~\eqref{eq:kernPotDegen2}. Since the natural amplitudes can only be defined for a two-electron system, the notion of relative phases only makes sense for two-electron systems. The relative phases therefore emphasize the special status of the interacting two-electron system concerning NO degeneracies.

All non-local one-body potentials in the kernel of the 1RDM response function have now been characterized for both the non-interacting case and the fully interacting Coulomb system. However, if one works in a small finite basis or uses some effective interaction which only affects some subspace, some special structure in the ground state might arise which causes additional potentials to be present in the kernel of the 1RDM response function. A complete proof including these cases would therefore require additional assumptions or a more extensive analysis which would depend on the specific details.

\subsection{Including degeneracies}

Now let us consider if additional potentials will be part of the kernel of the 1RDM response function if we allow for degenerate ground states. We will do this by first assuming that there exists some non-local one-body potential which only creates components in the degenerate subspace when acting on the initial state
\begin{align}\label{eq:UpotDef}
\hat{U}\ket{\Psi_0} = \sum_{i,j}u_{ji}\hat{\gamma}_{i,j}\ket{\Psi_0}
= \sum_{\mathclap{0 \leq K < D}}u^*_K\ket{\Psi_K} \neq 0 ,
\end{align}
where $u_K = \brakket{\Psi_0}{\hat{U}}{\Psi_K}$.
Subsequently we check whether the additional necessary condition~\eqref{eq:fDegenComm} is satisfied. This condition attains the following simple form in the NO basis
\begin{align}\label{eq:fDegen1RDM}
0 = \bigbrakket{\Psi_0}{\bigl[\hat{\gamma}_{kl},\hat{U}\bigr]}{\Psi_0} = (n_l - n_k)u_{kl} \qquad \forall_{k,l} .
\end{align}
This is a very interesting expression, since it tells us that only potentials $\hat{U}$ which have only non-zero matrix elements coupling degenerate NOs yield a zero 1RDM response.
This is a very stringent condition, especially in combination with the requirement that $\hat{U}\ket{\Psi_0}$ is only allowed to have components in the degenerate subspace of ground states.

Now let us investigate a number of systems of interest. First consider a system of non-interacting particles with degenerate ground states. These ground states can be constructed by first solving effective one-particle Schrödinger equations $\hat{h}\ket{\phi_i} = \epsilon_i\ket{\phi_i}$. Assuming that the orbital energies are ordered in increasing order, $\epsilon_i \leq \epsilon_{i+1}$, the complete span of degenerate ground states can be constructed by assembling all determinant with the lowest orbital energies
\begin{align*}
\ket{\Psi_{i_1,\dotsc,i_{N-k}}} = \crea{a}_{i_1}\dotsb\crea{a}_{i_{N-k}}\crea{a}_k\dotsb\crea{a}_1\ket{} ,
\end{align*}
where $i_1 < i_2 < \dotsb < i_{N-k} \in \mathcal{D} \isDefinedAs \{k+1,k+2,\dotsc,k+d\}$ and $d$ denotes the number of degenerate orbitals with orbital energy $\epsilon_{k+1} = \epsilon_{k+2} = \dotsb = \epsilon_{k+d}$. Note that this set of degenerate ground states is not unique. Arbitrary unitary transformations among the degenerate ground states yield different spans of the ground state subspace which are equally valid. In particular, the initial ground state of the response function can be chosen as any linear combination of the degenerate ground states.

A potential $\hat{U}$ satisfying condition~\eqref{eq:UpotDef} is readily constructed by setting $u_{ij} = 0$ if any $i,j \notin \mathcal{D}$ and choosing some $u_{i,j} \neq 0$ for both $i,j \in \mathcal{D}$. for the potential $\hat{U}$ to belong to the kernel of the 1RDM response function, we additionally need that $n_i = n_j$ for all elements $u_{i,j} \neq 0$. Note that the cases $n_i = n_j = 1$ and $n_i = n_j = 0$ were already covered before in the non-degenerate case, since $\ket{\Psi_0}$ will be actually an eigenstate of the potential $\hat{U}$. The treatment of degenerate non-interacting ground states extends this result to any potential coupling only degenerate NOs. So all potentials with $u_{ij} = 0$ for $n_i \neq n_j$ will belong to the kernel of the 1RDM response function in the non-interacting case.

Now let us consider a system with a spin degenerate ground states, $\ket{S,M}$. These states are eigenfunctions of the $\hat{S}_z$ operator, $\hat{S}_z\ket{S,M} = M\ket{S,M}$, so the $\hat{S}_z$ operator immediately belongs to kernel of the 1RDM response function. By operating with the $\hat{S}_{\pm}$ operators we can obtain other states in the degenerate subspace, $\hat{S}_{\pm}\ket{S,M} = C_{\pm}(S,M)\ket{S,M\pm1}$. In second quantizations, these raising and lowering operators can be expressed as
\begin{align*}
\hat{S}_+ &= \sum_k\crea{a}_{k\alpha}\anni{a}_{k\beta}	&
&\text{and}	&
\hat{S}_- &= \sum_k\crea{a}_{k\beta}\anni{a}_{k\alpha} .
\end{align*}
The operators $\hat{S}_{\pm}$ are not hermitian operators, but we can make two independent hermitian combinations which are properly hermitian
\begin{align*}
\hat{S}_x &= \half\bigl(\hat{S}_+ + \hat{S}_-\bigr)	&
&\text{and}	&
\hat{S}_y &= \frac{1}{2\I}\bigl(\hat{S}_+ - \hat{S}_-\bigr) .
\end{align*}
Since these operators only produce components in the degenerate subspace, we have found proper potentials $\hat{U}$ as in~\eqref{eq:UpotDef}. Now we need to check whether these operators satisfy~\eqref{eq:fDegen1RDM}. We see that the $\hat{S}_x$ and $\hat{S}_y$ operators couple the different spin components of each spatial orbital, so we need $n_{k\alpha} = n_{k\beta}$ for all $k$ for~\eqref{eq:fDegen1RDM} to hold. This degeneracy only occurs for $M = 0$, so only in the case that $\ket{S,0}$ is the initial state, the $\hat{S}_x$ and $\hat{S}_y$ operators also belong to the kernel of the 1RDM response function. Combined with our result for non-degenerate states, this means the $\hat{S}_z$ is always part of the the kernel of the 1RDM response function for Hamiltonians not depending on spin. If additionally the system is spin-compensated, i.e.\ $n_{k\alpha} = n_{k\beta}$ for all $k$, the $\hat{S}_x$ and $\hat{S}_y$ operators are also part of the kernel, irrespective if the ground state is degenerate or not. Note that the same considerations also hold for the angular momentum operators $\hat{\mat{L}}$ if the Hamiltonian is invariant under all rotations, e.g.\ atoms and the homogeneous electron gas, though we need to check for different degeneracies in the occupation spectrum. For example consider an atom. The $z$-axis can always be chosen such that the ground state is also an eigenstate of the $\hat{L}_z$ operator. For the $\hat{L}_x$ and $\hat{L}_y$ operators to be part of the kernel of the 1RDM response function as well, we need $n_{k,l,m} = n_{k,l,m'}$, which implies that the 1RDM will be unperturbed when we make rotations around an arbitrary axis.

\subsection{Ground 1RDM functional theory}
The generalized invertibility theorem for the non-degenerate case is equally valid for the time-\emph{in}dependent response function by taking the $s \to 0$ limit of the Laplace transformed quantities. The generalized invertibility theorem therefore provides the perfect opportunity to give a better classification of the uniqueness of the mapping from non-local one-body potentials to 1RDMs, $\hat{v} \mapsto \gamma$. As Gilbert already mentioned in 1975~\cite{Gilbert1975}, the class of potentials which map to the same ground state 1RDM will be larger than in DFT, but to the author's knowledge no attempt has been made to give a full classification of this non-uniqueness. We will show that the kernel of the time-dependent 1RDM response function exactly corresponds to the non-uniqueness of the non-local potential in ground 1RDM functional theory in the non-degenerate case.

As Gilbert already showed~\cite{Gilbert1975}, the second part of the Hohenberg--Kohn theorem can straightforwardly be generalized to 1RDMs: the 1RDM of a non-degenerate ground state is unique. In other words, consider all the ground states corresponding to different non-local potentials, then there is a one-to-one correspondence between the non-degenerate ground states and their corresponding 1RDMs.

Now assume that there are two (non-local) potentials, $\hat{v}_1$ and $\hat{v}_2$, yielding the same non-degenerate ground state. Since the Schrödinger equation is linear, the potentials $\hat{v}_{\lambda} = (1-\lambda)\hat{v}_1 + \lambda\hat{v}_2$ yield exactly the same non-degenerate ground state. The set of potentials which yield the same non-degenerate ground state is therefore (simply) connected. To determine this set, it is therefore sufficient to consider a perturbation to one of these potentials and check which potentials do not lead to a response to any order. As we have shown before, the first order 1RDM response only vanishes if the ground state is an eigenstate of the perturbation, but this also immediately implies that the response will vanish to any order. We can therefore conclude that the kernel of the 1RDM response function exactly coincides with the class of potentials yielding the same ground state 1RDM. More precisely, two non-local one-body potentials yield the same ground state (1RDM) if and only if their difference is part of the kernel of the 1RDM response function. Note that these considerations are not special for the 1RDM, but can be applied to any density-functional-like theory for which we are able to characterize the kernel of the response function.

\begin{figure}[t]
  \includegraphics[width=\columnwidth]{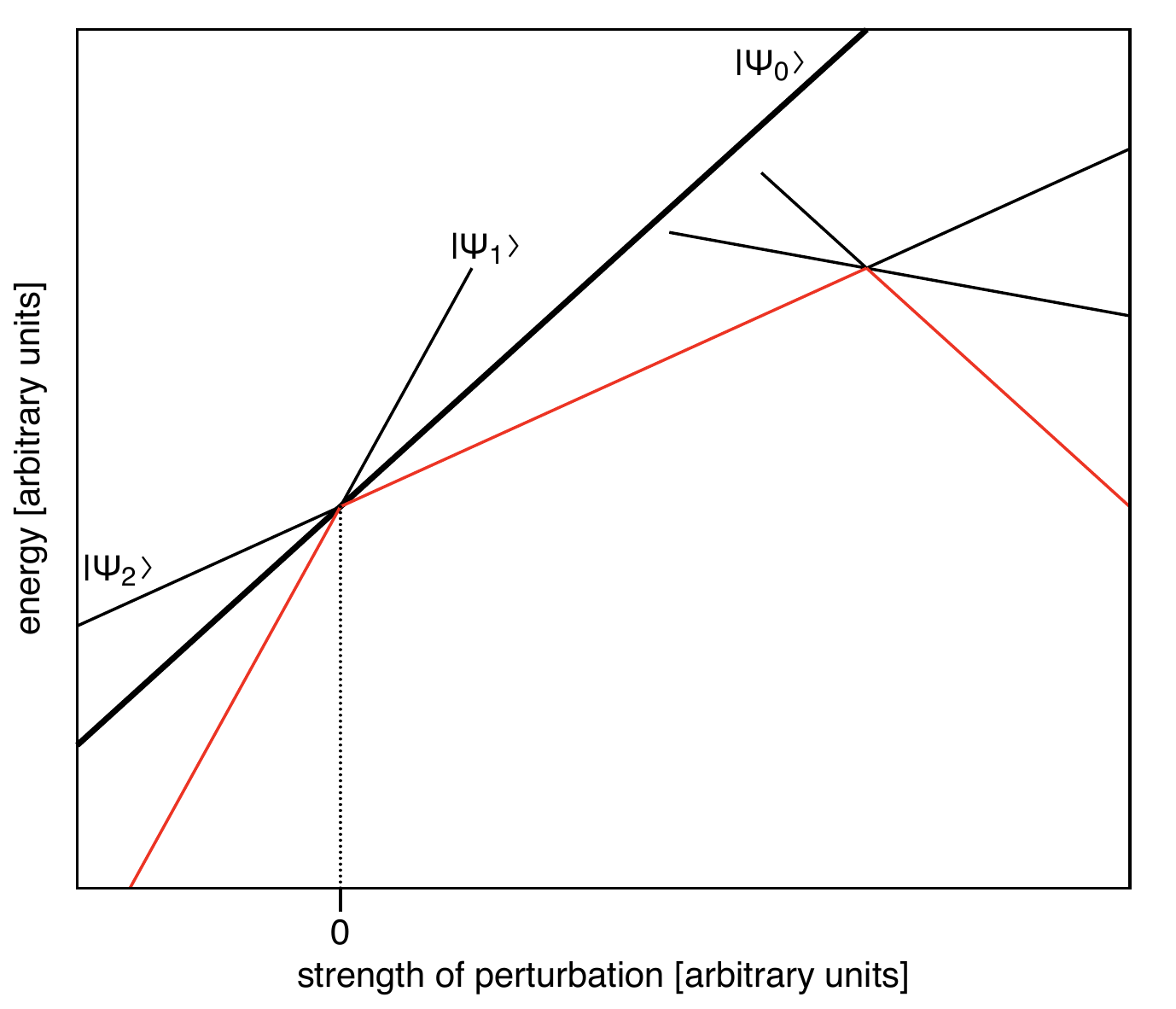}
  \caption{A sketch of the dependence of the lowest eigenstates of some quantum system as a function of a perturbation. The energy dependence of the initial state in the time-dependent case is shown by the thick line and is simply a linear function. Time-\emph{in}dependent perturbation theory always selects the lowest perturbed state, so $\ket{\Psi_1}$ on the negative side and $\ket{\Psi_2}$ on the positive site up to the next degeneracy point. To emphasize the jumps at the degeneracy points, the lowest energies are colored red.}
  \label{fig:Degeneracy}
\end{figure}

The degenerate case is beyond the scope of this article. The main reason is that the degenerate case is handled in a fundamentally different manner in time-dependent and time-independent perturbation theory. Time-independent perturbation theory is based on the time-independent Schrödinger equation, which is an eigenvalue equation. The ground state is therefore only specified up to its degenerate subspace, from which an appropriate $\ket{\Psi_0}$ needs to be chosen. This is reflected in its perturbation theory, since we need to diagonalize the perturbation in the degenerate subspace and take the lowest eigenvalue. The perturbed state therefore depends on the direction of the perturbation as illustrated by the lowest energy surface in Fig.~\ref{fig:Degeneracy}. Time-dependent perturbation theory is based on the time-dependent Schrödinger equation, which is an initial value problem. The initial state $\ket{\Psi_0}$ is therefore completely specified from the start, even in the degenerate case. Therefore, contrary to the time-independent case, we do no need to (and can not) diagonalize the perturbation in the degenerate subspace to select an appropriate zeroth order state, since it is simply specified from the start as $\ket{\Psi_0}$. The time-dependent situation is illustrated in Fig.~\ref{fig:Degeneracy} by the thick line. Due to the fundamental difference in dealing with degeneracies, the result for the time-dependent response function does not straightforwardly carry over to the time-independent response function in the degenerate case and a separate treatment is required.

\section{Conclusion}
To summarize, I have generalized the first step of the invertibility theorem for the density response function by Van Leeuwen~\cite{Leeuwen2001}, to arbitrary operators and to degenerate ground states. For the nontrivial absence of response, it is sufficient that initial ground state is an eigenstate of the perturbation operator and also necessary in the case of a non-degenerate ground state. For a degenerate ground state, however, the action of the perturbation operator is allowed to yield additional components in the degenerate subspace, though the expectation value of the commutator of the perturbation with any operator under consideration needs to vanish to yield zero response as an additional condition~\eqref{eq:fDegenComm}.

The theorem can be used to establish density-functional-like theories in the time-dependent linear response regime. The restriction to ground states is not very severe, since this is the initial state which is used almost exclusively in practical linear response calculations. The determination of the kernel of the time-dependent response function also immediately carries over to the time-\emph{in}dependent response function if the initial ground state is non-degenerate. This result is useful to establish time-independent density-functional-like theories. The kernel of the response function exactly coincides with the $v \mapsto Q$ mapping in the non-degenerate case. The non-uniqueness of the potential can therefore be fully characterized as in the first Hohenberg--Kohn theorem for DFT.

The generalized invertibility theorem has been applied to the density response function and it has been established that only the spatially constant potential belongs to its kernel, even for a degenerate ground state. Applying the theorem to the 1RDM response function revealed that not only the constant time-dependent shift is part of the kernel, but also generators of continuous symmetries are possibly included. For non-relativistic Hamiltonians this would always be the $\hat{S}_z$ operator and if the NOs are degenerate in both spin channels, $n_{k\alpha} = n_{k\beta}$, also the other components of $\hat{\mat{S}}$ belong to the kernel of the 1RDM response function, cf.\ non-collinear spin-DFT~\cite{BarthHedin1972, EschrigPickett2001}. Also the angular momentum operators are possibly included in the kernel of the 1RDM response function if the Hamiltonian is invariant under the corresponding rotations. The additional condition~\eqref{eq:fDegenComm} requires the relevant NOs to be degenerate as well. It is obvious that when spin-orbit coupling is included, the relevant operators to be considered would be $\hat{\mat{J}} \isDefinedAs \hat{\mat{L}} + \hat{\mat{S}}$ instead of $\hat{\mat{L}}$ and $\hat{\mat{S}}$ separately. For homogeneous systems, e.g.\ the homogeneous electron gas, also the momentum operators $-\I\nabla$ need to be considered.
Further, the matrix elements of the non-local potential which couple within the fully unoccupied block or within the fully occupied block also do not lead to a first order response. 
In the non-interacting case actually all potentials coupling only degenerate NOs belong to the kernel of the 1RDM response function.
Due to the intimate relation between a two-electron state and its 1RDM, degeneracies of the natural occupation numbers give rise to additional non-local potentials in the kernel of the 1RDM response function, whose matrix elements couple the natural orbitals within the degenerate sub-block.
This result not only puts time-dependent linear response 1RDM functional theory on a rigorous basis but is also of high importance for ground state 1RDM functional theory, because it allows for a full characterization of the non-uniqueness of the non-local potential for non-degenerate ground states for the first time.

\begin{acknowledgments}
The author would like to thank dr.\ M.~Ruggenthaler prof.dr.\ R.~van Leeuwen and prof.dr.\ E.J. Baerends for stimulating discussions and prof.dr.\ E.K.U. Gross for pointing out that the generalized invertibility theorem also solves the non-uniqueness problem in 1RDM functional theory. Also the critical remarks by the first reviewer are much appreciated.
Support from the Netherlands Foundation for Research NWO (722.012.013) through a VENI grant is gratefully acknowledged.
\end{acknowledgments}

\nocite{PenzRuggenthaler2015}

\bibliography{bible}

\begin{thebibliography}{61}%
\makeatletter
\providecommand \@ifxundefined [1]{%
 \@ifx{#1\undefined}
}%
\providecommand \@ifnum [1]{%
 \ifnum #1\expandafter \@firstoftwo
 \else \expandafter \@secondoftwo
 \fi
}%
\providecommand \@ifx [1]{%
 \ifx #1\expandafter \@firstoftwo
 \else \expandafter \@secondoftwo
 \fi
}%
\providecommand \natexlab [1]{#1}%
\providecommand \enquote  [1]{``#1''}%
\providecommand \bibnamefont  [1]{#1}%
\providecommand \bibfnamefont [1]{#1}%
\providecommand \citenamefont [1]{#1}%
\providecommand \href@noop [0]{\@secondoftwo}%
\providecommand \href [0]{\begingroup \@sanitize@url \@href}%
\providecommand \@href[1]{\@@startlink{#1}\@@href}%
\providecommand \@@href[1]{\endgroup#1\@@endlink}%
\providecommand \@sanitize@url [0]{\catcode `\\12\catcode `\$12\catcode
  `\&12\catcode `\#12\catcode `\^12\catcode `\_12\catcode `\%12\relax}%
\providecommand \@@startlink[1]{}%
\providecommand \@@endlink[0]{}%
\providecommand \url  [0]{\begingroup\@sanitize@url \@url }%
\providecommand \@url [1]{\endgroup\@href {#1}{\urlprefix }}%
\providecommand \urlprefix  [0]{URL }%
\providecommand \Eprint [0]{\href }%
\providecommand \doibase [0]{http://dx.doi.org/}%
\providecommand \selectlanguage [0]{\@gobble}%
\providecommand \bibinfo  [0]{\@secondoftwo}%
\providecommand \bibfield  [0]{\@secondoftwo}%
\providecommand \translation [1]{[#1]}%
\providecommand \BibitemOpen [0]{}%
\providecommand \bibitemStop [0]{}%
\providecommand \bibitemNoStop [0]{.\EOS\space}%
\providecommand \EOS [0]{\spacefactor3000\relax}%
\providecommand \BibitemShut  [1]{\csname bibitem#1\endcsname}%
\let\auto@bib@innerbib\@empty
\bibitem [{\citenamefont {Runge}\ and\ \citenamefont
  {Gross}(1984)}]{RungeGross1984}%
  \BibitemOpen
  \bibfield  {author} {\bibinfo {author} {\bibfnamefont {E.}~\bibnamefont
  {Runge}}\ and\ \bibinfo {author} {\bibfnamefont {E.~K.~U.}\ \bibnamefont
  {Gross}},\ }\href {\doibase 10.1103/PhysRevLett.52.997} {\bibfield  {journal}
  {\bibinfo  {journal} {Phys. Rev. Lett.}\ }\textbf {\bibinfo {volume} {52}},\
  \bibinfo {pages} {997} (\bibinfo {year} {1984})}\BibitemShut {NoStop}%
\bibitem [{\citenamefont {Marques}\ \emph {et~al.}(2006)\citenamefont
  {Marques}, \citenamefont {Ulrich}, \citenamefont {Nogueira}, \citenamefont
  {Rubio}, \citenamefont {Burke},\ and\ \citenamefont {Gross}}]{tddft2006}%
  \BibitemOpen
  \bibinfo {editor} {\bibfnamefont {M.~A.~L.}\ \bibnamefont {Marques}},
  \bibinfo {editor} {\bibfnamefont {C.~A.}\ \bibnamefont {Ulrich}}, \bibinfo
  {editor} {\bibfnamefont {F.}~\bibnamefont {Nogueira}}, \bibinfo {editor}
  {\bibfnamefont {A.}~\bibnamefont {Rubio}}, \bibinfo {editor} {\bibfnamefont
  {K.}~\bibnamefont {Burke}}, \ and\ \bibinfo {editor} {\bibfnamefont
  {E.~K.~U.}\ \bibnamefont {Gross}},\ eds.,\ \href@noop {} {\emph {\bibinfo
  {title} {Time-Dependent Density Functional Theory}}},\ \bibinfo {series}
  {Lect. Notes. Phys.}\ No.\ \bibinfo {number} {706}\ (\bibinfo  {publisher}
  {Springer-Verlag},\ \bibinfo {address} {Berlin Heidelberg},\ \bibinfo {year}
  {2006})\BibitemShut {NoStop}%
\bibitem [{\citenamefont {Casida}\ and\ \citenamefont
  {Huix-Rotllant}(2012)}]{CasidaHuix-Rotllant2012}%
  \BibitemOpen
  \bibfield  {author} {\bibinfo {author} {\bibfnamefont {M.~E.}\ \bibnamefont
  {Casida}}\ and\ \bibinfo {author} {\bibfnamefont {M.}~\bibnamefont
  {Huix-Rotllant}},\ }\href {\doibase 10.1146/annurev-physchem-032511-143803}
  {\bibfield  {journal} {\bibinfo  {journal} {Annu. Rev. Phys. Chem.}\ }\textbf
  {\bibinfo {volume} {63}},\ \bibinfo {pages} {287} (\bibinfo {year}
  {2012})}\BibitemShut {NoStop}%
\bibitem [{\citenamefont {Ullrich}(2012)}]{Ullrich2012}%
  \BibitemOpen
  \bibfield  {author} {\bibinfo {author} {\bibfnamefont {C.~A.}\ \bibnamefont
  {Ullrich}},\ }\href@noop {} {{\selectlanguage {english}\emph {\bibinfo
  {title} {Time-Dependent Density-Functional Theory: Concepts and
  Applications}}}},\ \bibinfo {edition} {1st}\ ed.,\ Oxford Graduate Texts\
  (\bibinfo  {publisher} {Oxford University Press},\ \bibinfo {address} {New
  York},\ \bibinfo {year} {2012})\BibitemShut {NoStop}%
\bibitem [{\citenamefont {Marques}\ \emph {et~al.}(2012)\citenamefont
  {Marques}, \citenamefont {Maitra}, \citenamefont {Nogueira}, \citenamefont
  {Gross},\ and\ \citenamefont {Rubio}}]{tddft2012}%
  \BibitemOpen
  \bibinfo {editor} {\bibfnamefont {M.~A.~L.}\ \bibnamefont {Marques}},
  \bibinfo {editor} {\bibfnamefont {N.~T.}\ \bibnamefont {Maitra}}, \bibinfo
  {editor} {\bibfnamefont {F.~M.~S.}\ \bibnamefont {Nogueira}}, \bibinfo
  {editor} {\bibfnamefont {E.~K.~U.}\ \bibnamefont {Gross}}, \ and\ \bibinfo
  {editor} {\bibfnamefont {A.}~\bibnamefont {Rubio}},\ eds.,\ \href {\doibase
  10.1007/978-3-642-23518-4} {\emph {\bibinfo {title} {Fundamentals of
  Time-Dependent Density Functional Theory}}},\ \bibinfo {edition} {1st}\ ed.,\
  \bibinfo {series} {Lecture Notes in Physics}, Vol.\ \bibinfo {volume} {837}\
  (\bibinfo  {publisher} {Springer-Verlag},\ \bibinfo {address} {Berlin
  Heidelberg},\ \bibinfo {year} {2012})\BibitemShut {NoStop}%
\bibitem [{\citenamefont {van Leeuwen}(2001)}]{Leeuwen2001}%
  \BibitemOpen
  \bibfield  {author} {\bibinfo {author} {\bibfnamefont {R.}~\bibnamefont {van
  Leeuwen}},\ }\href@noop {} {\bibfield  {journal} {\bibinfo  {journal} {Int.
  J. Mod. Phys. B}\ }\textbf {\bibinfo {volume} {15}},\ \bibinfo {pages} {1969}
  (\bibinfo {year} {2001})}\BibitemShut {NoStop}%
\bibitem [{\citenamefont {Tokatly}(2011{\natexlab{a}})}]{Tokatly2011a}%
  \BibitemOpen
  \bibfield  {author} {\bibinfo {author} {\bibfnamefont {I.~V.}\ \bibnamefont
  {Tokatly}},\ }\href {\doibase 10.1103/PhysRevB.83.035127} {\bibfield
  {journal} {\bibinfo  {journal} {Phys. Rev. B}\ }\textbf {\bibinfo {volume}
  {83}},\ \bibinfo {pages} {035127} (\bibinfo {year}
  {2011}{\natexlab{a}})}\BibitemShut {NoStop}%
\bibitem [{\citenamefont {Tokatly}(2011{\natexlab{b}})}]{Tokatly2011b}%
  \BibitemOpen
  \bibfield  {author} {\bibinfo {author} {\bibfnamefont {I.~V.}\ \bibnamefont
  {Tokatly}},\ }\href {\doibase 10.1016/j.chemphys.2011.04.005} {\bibfield
  {journal} {\bibinfo  {journal} {Chem. Phys.}\ }\textbf {\bibinfo {volume}
  {391}},\ \bibinfo {pages} {78} (\bibinfo {year}
  {2011}{\natexlab{b}})}\BibitemShut {NoStop}%
\bibitem [{\citenamefont {Farzanehpour}\ and\ \citenamefont
  {Tokatly}(2012)}]{FarzanehpourTokatly2012}%
  \BibitemOpen
  \bibfield  {author} {\bibinfo {author} {\bibfnamefont {M.}~\bibnamefont
  {Farzanehpour}}\ and\ \bibinfo {author} {\bibfnamefont {I.~V.}\ \bibnamefont
  {Tokatly}},\ }\href {\doibase 10.1103/PhysRevB.86.125130} {\bibfield
  {journal} {\bibinfo  {journal} {Phys. Rev. B}\ }\textbf {\bibinfo {volume}
  {86}},\ \bibinfo {pages} {125130} (\bibinfo {year} {2012})}\BibitemShut
  {NoStop}%
\bibitem [{\citenamefont {Ruggenthaler}\ and\ \citenamefont {van
  Leeuwen}(2011)}]{RuggenthalerLeeuwen2011}%
  \BibitemOpen
  \bibfield  {author} {\bibinfo {author} {\bibfnamefont {M.}~\bibnamefont
  {Ruggenthaler}}\ and\ \bibinfo {author} {\bibfnamefont {R.}~\bibnamefont {van
  Leeuwen}},\ }\href {\doibase 10.1209/0295-5075/95/13001} {\bibfield
  {journal} {\bibinfo  {journal} {Europhys. Lett.}\ }\textbf {\bibinfo {volume}
  {95}},\ \bibinfo {pages} {13001} (\bibinfo {year} {2011})}\BibitemShut
  {NoStop}%
\bibitem [{\citenamefont {Ruggenthaler}\ \emph {et~al.}(2012)\citenamefont
  {Ruggenthaler}, \citenamefont {Giesbertz}, \citenamefont {Penz},\ and\
  \citenamefont {van Leeuwen}}]{RuggenthalerGiesbertzPenz2012}%
  \BibitemOpen
  \bibfield  {author} {\bibinfo {author} {\bibfnamefont {M.}~\bibnamefont
  {Ruggenthaler}}, \bibinfo {author} {\bibfnamefont {K.~J.~H.}\ \bibnamefont
  {Giesbertz}}, \bibinfo {author} {\bibfnamefont {M.}~\bibnamefont {Penz}}, \
  and\ \bibinfo {author} {\bibfnamefont {R.}~\bibnamefont {van Leeuwen}},\
  }\href {\doibase 10.1103/PhysRevA.85.052504} {\bibfield  {journal} {\bibinfo
  {journal} {Phys. Rev. A}\ }\textbf {\bibinfo {volume} {85}},\ \bibinfo
  {pages} {052504} (\bibinfo {year} {2012})}\BibitemShut {NoStop}%
\bibitem [{\citenamefont {Ruggenthaler}\ \emph {et~al.}(2015)\citenamefont
  {Ruggenthaler}, \citenamefont {Penz},\ and\ \citenamefont {van
  Leeuwen}}]{RuggenthalerPenzLeeuwen2015}%
  \BibitemOpen
  \bibfield  {author} {\bibinfo {author} {\bibfnamefont {M.}~\bibnamefont
  {Ruggenthaler}}, \bibinfo {author} {\bibfnamefont {M.}~\bibnamefont {Penz}},
  \ and\ \bibinfo {author} {\bibfnamefont {R.}~\bibnamefont {van Leeuwen}},\
  }\href {\doibase 10.1088/0953-8984/27/20/203202} {\bibfield  {journal}
  {\bibinfo  {journal} {J. Phys.: Condens. Matter}\ }\textbf {\bibinfo {volume}
  {27}},\ \bibinfo {pages} {203202} (\bibinfo {year} {2015})}\BibitemShut
  {NoStop}%
\bibitem [{\citenamefont {Kohn}\ and\ \citenamefont
  {Sham}(1965)}]{KohnSham1965}%
  \BibitemOpen
  \bibfield  {author} {\bibinfo {author} {\bibfnamefont {W.}~\bibnamefont
  {Kohn}}\ and\ \bibinfo {author} {\bibfnamefont {L.~J.}\ \bibnamefont
  {Sham}},\ }\href {\doibase 10.1103/PhysRev.140.A1133} {\bibfield  {journal}
  {\bibinfo  {journal} {Phys. Rev.}\ }\textbf {\bibinfo {volume} {140}},\
  \bibinfo {pages} {A1133} (\bibinfo {year} {1965})}\BibitemShut {NoStop}%
\bibitem [{\citenamefont {Ruggenthaler}\ \emph {et~al.}(2013)\citenamefont
  {Ruggenthaler}, \citenamefont {Nielsen},\ and\ \citenamefont {van
  Leeuwen}}]{RuggenthalerNielsenLeeuwen2013}%
  \BibitemOpen
  \bibfield  {author} {\bibinfo {author} {\bibfnamefont {M.}~\bibnamefont
  {Ruggenthaler}}, \bibinfo {author} {\bibfnamefont {S.~E.~B.}\ \bibnamefont
  {Nielsen}}, \ and\ \bibinfo {author} {\bibfnamefont {R.}~\bibnamefont {van
  Leeuwen}},\ }\href {\doibase 10.1103/PhysRevA.88.022512} {\bibfield
  {journal} {\bibinfo  {journal} {Phys. Rev. A}\ }\textbf {\bibinfo {volume}
  {88}},\ \bibinfo {pages} {022512} (\bibinfo {year} {2013})}\BibitemShut
  {NoStop}%
\bibitem [{\citenamefont {van Gisbergen}\ \emph {et~al.}(1998)\citenamefont
  {van Gisbergen}, \citenamefont {Kootstra}, \citenamefont {Schipper},
  \citenamefont {Gritsenko}, \citenamefont {Snijders},\ and\ \citenamefont
  {Baerends}}]{GisbergenKootstraSchipper1998}%
  \BibitemOpen
  \bibfield  {author} {\bibinfo {author} {\bibfnamefont {S.~J.~A.}\
  \bibnamefont {van Gisbergen}}, \bibinfo {author} {\bibfnamefont
  {F.}~\bibnamefont {Kootstra}}, \bibinfo {author} {\bibfnamefont {P.~R.~T.}\
  \bibnamefont {Schipper}}, \bibinfo {author} {\bibfnamefont {O.~V.}\
  \bibnamefont {Gritsenko}}, \bibinfo {author} {\bibfnamefont {J.~G.}\
  \bibnamefont {Snijders}}, \ and\ \bibinfo {author} {\bibfnamefont {E.~J.}\
  \bibnamefont {Baerends}},\ }\href {\doibase 10.1103/PhysRevA.57.2556}
  {\bibfield  {journal} {\bibinfo  {journal} {Phys. Rev. A}\ }\textbf {\bibinfo
  {volume} {57}},\ \bibinfo {pages} {2556} (\bibinfo {year}
  {1998})}\BibitemShut {NoStop}%
\bibitem [{\citenamefont {Schipper}\ \emph {et~al.}(2000)\citenamefont
  {Schipper}, \citenamefont {Gritsenko}, \citenamefont {van Gisbergen},\ and\
  \citenamefont {Baerends}}]{SchipperGritsenkoGisbergen2000}%
  \BibitemOpen
  \bibfield  {author} {\bibinfo {author} {\bibfnamefont {P.~R.~T.}\
  \bibnamefont {Schipper}}, \bibinfo {author} {\bibfnamefont {O.~V.}\
  \bibnamefont {Gritsenko}}, \bibinfo {author} {\bibfnamefont {S.~J.~A.}\
  \bibnamefont {van Gisbergen}}, \ and\ \bibinfo {author} {\bibfnamefont
  {E.~J.}\ \bibnamefont {Baerends}},\ }\href {\doibase 10.1063/1.480688}
  {\bibfield  {journal} {\bibinfo  {journal} {J. Chem. Phys.}\ }\textbf
  {\bibinfo {volume} {112}},\ \bibinfo {pages} {1344} (\bibinfo {year}
  {2000})}\BibitemShut {NoStop}%
\bibitem [{\citenamefont {Appel}\ \emph {et~al.}(2003)\citenamefont {Appel},
  \citenamefont {Gross},\ and\ \citenamefont {Burke}}]{AppelGrossBurke2003}%
  \BibitemOpen
  \bibfield  {author} {\bibinfo {author} {\bibfnamefont {H.}~\bibnamefont
  {Appel}}, \bibinfo {author} {\bibfnamefont {E.~K.~U.}\ \bibnamefont {Gross}},
  \ and\ \bibinfo {author} {\bibfnamefont {K.}~\bibnamefont {Burke}},\ }\href
  {\doibase 10.1103/PhysRevLett.90.043005} {\bibfield  {journal} {\bibinfo
  {journal} {Phys. Rev. Lett.}\ }\textbf {\bibinfo {volume} {90}},\ \bibinfo
  {pages} {043005} (\bibinfo {year} {2003})}\BibitemShut {NoStop}%
\bibitem [{\citenamefont {Dreuw}\ \emph {et~al.}(2003)\citenamefont {Dreuw},
  \citenamefont {Weisman},\ and\ \citenamefont
  {Head-Gordon}}]{DreuwWeismanHead-Gordon2003}%
  \BibitemOpen
  \bibfield  {author} {\bibinfo {author} {\bibfnamefont {A.}~\bibnamefont
  {Dreuw}}, \bibinfo {author} {\bibfnamefont {J.~L.}\ \bibnamefont {Weisman}},
  \ and\ \bibinfo {author} {\bibfnamefont {M.}~\bibnamefont {Head-Gordon}},\
  }\href {\doibase 10.1063/1.1590951} {\bibfield  {journal} {\bibinfo
  {journal} {J. Chem. Phys.}\ }\textbf {\bibinfo {volume} {119}},\ \bibinfo
  {pages} {2943} (\bibinfo {year} {2003})}\BibitemShut {NoStop}%
\bibitem [{\citenamefont {Gritsenko}\ and\ \citenamefont
  {Baerends}(2004)}]{GritsenkoBaerends2004b}%
  \BibitemOpen
  \bibfield  {author} {\bibinfo {author} {\bibfnamefont {O.}~\bibnamefont
  {Gritsenko}}\ and\ \bibinfo {author} {\bibfnamefont {E.~J.}\ \bibnamefont
  {Baerends}},\ }\href {\doibase 10.1063/1.1759320} {\bibfield  {journal}
  {\bibinfo  {journal} {J. Chem. Phys.}\ }\textbf {\bibinfo {volume} {121}},\
  \bibinfo {pages} {655} (\bibinfo {year} {2004})}\BibitemShut {NoStop}%
\bibitem [{\citenamefont {Rohlfing}\ and\ \citenamefont
  {Louie}(1998)}]{RohlfingLouie1998}%
  \BibitemOpen
  \bibfield  {author} {\bibinfo {author} {\bibfnamefont {M.}~\bibnamefont
  {Rohlfing}}\ and\ \bibinfo {author} {\bibfnamefont {S.~G.}\ \bibnamefont
  {Louie}},\ }\href {\doibase 10.1103/PhysRevLett.81.2312} {\bibfield
  {journal} {\bibinfo  {journal} {Phys. Rev. Lett.}\ }\textbf {\bibinfo
  {volume} {81}},\ \bibinfo {pages} {2312} (\bibinfo {year}
  {1998})}\BibitemShut {NoStop}%
\bibitem [{\citenamefont {Benedict}\ \emph {et~al.}(1998)\citenamefont
  {Benedict}, \citenamefont {Shirley},\ and\ \citenamefont
  {Bohn}}]{BenedictShirleyBohn1998}%
  \BibitemOpen
  \bibfield  {author} {\bibinfo {author} {\bibfnamefont {L.~X.}\ \bibnamefont
  {Benedict}}, \bibinfo {author} {\bibfnamefont {E.~L.}\ \bibnamefont
  {Shirley}}, \ and\ \bibinfo {author} {\bibfnamefont {R.~B.}\ \bibnamefont
  {Bohn}},\ }\href {\doibase 10.1103/PhysRevLett.80.4514} {\bibfield  {journal}
  {\bibinfo  {journal} {Phys. Rev. Lett.}\ }\textbf {\bibinfo {volume} {80}},\
  \bibinfo {pages} {4514} (\bibinfo {year} {1998})}\BibitemShut {NoStop}%
\bibitem [{\citenamefont {Baerends}\ \emph {et~al.}(2013)\citenamefont
  {Baerends}, \citenamefont {Gritsenko},\ and\ \citenamefont {van
  Meer}}]{BaerendsGritsenkoMeer2013}%
  \BibitemOpen
  \bibfield  {author} {\bibinfo {author} {\bibfnamefont {E.~J.}\ \bibnamefont
  {Baerends}}, \bibinfo {author} {\bibfnamefont {O.~V.}\ \bibnamefont
  {Gritsenko}}, \ and\ \bibinfo {author} {\bibfnamefont {R.}~\bibnamefont {van
  Meer}},\ }\href {\doibase 10.1039/C3CP52547C} {\bibfield  {journal} {\bibinfo
   {journal} {Phys. Chem. Chem. Phys.}\ }\textbf {\bibinfo {volume} {15}},\
  \bibinfo {pages} {16408} (\bibinfo {year} {2013})}\BibitemShut {NoStop}%
\bibitem [{\citenamefont {Reining}\ \emph {et~al.}(2002)\citenamefont
  {Reining}, \citenamefont {Olevano}, \citenamefont {Rubio},\ and\
  \citenamefont {Onida}}]{ReiningOlevanoRubio2002}%
  \BibitemOpen
  \bibfield  {author} {\bibinfo {author} {\bibfnamefont {L.}~\bibnamefont
  {Reining}}, \bibinfo {author} {\bibfnamefont {V.}~\bibnamefont {Olevano}},
  \bibinfo {author} {\bibfnamefont {A.}~\bibnamefont {Rubio}}, \ and\ \bibinfo
  {author} {\bibfnamefont {G.}~\bibnamefont {Onida}},\ }\href {\doibase
  10.1103/PhysRevLett.88.066404} {\bibfield  {journal} {\bibinfo  {journal}
  {Phys. Rev. Lett.}\ }\textbf {\bibinfo {volume} {88}},\ \bibinfo {pages}
  {066404} (\bibinfo {year} {2002})}\BibitemShut {NoStop}%
\bibitem [{\citenamefont {Yanai}\ \emph {et~al.}(2004)\citenamefont {Yanai},
  \citenamefont {Tew},\ and\ \citenamefont {Handy}}]{YanaiTewHandy2004}%
  \BibitemOpen
  \bibfield  {author} {\bibinfo {author} {\bibfnamefont {T.}~\bibnamefont
  {Yanai}}, \bibinfo {author} {\bibfnamefont {D.~P.}\ \bibnamefont {Tew}}, \
  and\ \bibinfo {author} {\bibfnamefont {N.~C.}\ \bibnamefont {Handy}},\ }\href
  {\doibase 10.1016/j.cplett.2004.06.011} {\bibfield  {journal} {\bibinfo
  {journal} {Chem. Phys. Lett.}\ }\textbf {\bibinfo {volume} {393}},\ \bibinfo
  {pages} {51} (\bibinfo {year} {2004})}\BibitemShut {NoStop}%
\bibitem [{\citenamefont {Yang}\ and\ \citenamefont
  {Ullrich}(2013)}]{YangUllrich2013}%
  \BibitemOpen
  \bibfield  {author} {\bibinfo {author} {\bibfnamefont {Z.-h.}\ \bibnamefont
  {Yang}}\ and\ \bibinfo {author} {\bibfnamefont {C.~A.}\ \bibnamefont
  {Ullrich}},\ }\href {\doibase 10.1103/PhysRevB.87.195204} {\bibfield
  {journal} {\bibinfo  {journal} {Phys. Rev. B}\ }\textbf {\bibinfo {volume}
  {87}},\ \bibinfo {pages} {195204} (\bibinfo {year} {2013})}\BibitemShut
  {NoStop}%
\bibitem [{\citenamefont {Maitra}\ \emph {et~al.}(2004)\citenamefont {Maitra},
  \citenamefont {Zhang}, \citenamefont {Cave},\ and\ \citenamefont
  {Burke}}]{MaitraZhangCave2004}%
  \BibitemOpen
  \bibfield  {author} {\bibinfo {author} {\bibfnamefont {N.~T.}\ \bibnamefont
  {Maitra}}, \bibinfo {author} {\bibfnamefont {F.}~\bibnamefont {Zhang}},
  \bibinfo {author} {\bibfnamefont {R.~J.}\ \bibnamefont {Cave}}, \ and\
  \bibinfo {author} {\bibfnamefont {K.}~\bibnamefont {Burke}},\ }\href
  {\doibase 10.1063/1.1651060} {\bibfield  {journal} {\bibinfo  {journal} {J.
  Chem. Phys.}\ }\textbf {\bibinfo {volume} {120}},\ \bibinfo {pages} {5932}
  (\bibinfo {year} {2004})}\BibitemShut {NoStop}%
\bibitem [{\citenamefont {Cave}\ \emph {et~al.}(2004)\citenamefont {Cave},
  \citenamefont {Zhang}, \citenamefont {Maitra},\ and\ \citenamefont
  {Burke}}]{CaveZhangMaitra2004}%
  \BibitemOpen
  \bibfield  {author} {\bibinfo {author} {\bibfnamefont {R.~J.}\ \bibnamefont
  {Cave}}, \bibinfo {author} {\bibfnamefont {F.}~\bibnamefont {Zhang}},
  \bibinfo {author} {\bibfnamefont {N.~T.}\ \bibnamefont {Maitra}}, \ and\
  \bibinfo {author} {\bibfnamefont {K.}~\bibnamefont {Burke}},\ }\href
  {\doibase 10.1016/j.cplett.2004.03.051} {\bibfield  {journal} {\bibinfo
  {journal} {Chem. Phys. Lett.}\ }\textbf {\bibinfo {volume} {389}},\ \bibinfo
  {pages} {39} (\bibinfo {year} {2004})}\BibitemShut {NoStop}%
\bibitem [{\citenamefont {Gritsenko}\ \emph {et~al.}(2000)\citenamefont
  {Gritsenko}, \citenamefont {van Gisbergen}, \citenamefont {G{\"o}rling},\
  and\ \citenamefont {Baerends}}]{GritsenkoGisbergenGorling2000}%
  \BibitemOpen
  \bibfield  {author} {\bibinfo {author} {\bibfnamefont {O.}~\bibnamefont
  {Gritsenko}}, \bibinfo {author} {\bibfnamefont {S.~J.~A.}\ \bibnamefont {van
  Gisbergen}}, \bibinfo {author} {\bibfnamefont {A.}~\bibnamefont
  {G{\"o}rling}}, \ and\ \bibinfo {author} {\bibfnamefont {E.~J.}\ \bibnamefont
  {Baerends}},\ }\href {\doibase 10.1063/1.1318750} {\bibfield  {journal}
  {\bibinfo  {journal} {J. Chem. Phys.}\ }\textbf {\bibinfo {volume} {113}},\
  \bibinfo {pages} {8478} (\bibinfo {year} {2000})}\BibitemShut {NoStop}%
\bibitem [{\citenamefont {Giesbertz}\ and\ \citenamefont
  {Baerends}(2008)}]{GiesbertzBaerends2008}%
  \BibitemOpen
  \bibfield  {author} {\bibinfo {author} {\bibfnamefont {K.~J.~H.}\
  \bibnamefont {Giesbertz}}\ and\ \bibinfo {author} {\bibfnamefont {E.~J.}\
  \bibnamefont {Baerends}},\ }\href {\doibase 10.1016/j.cplett.2008.07.018}
  {\bibfield  {journal} {\bibinfo  {journal} {Chem. Phys. Lett.}\ }\textbf
  {\bibinfo {volume} {461}},\ \bibinfo {pages} {338} (\bibinfo {year}
  {2008})}\BibitemShut {NoStop}%
\bibitem [{\citenamefont {Giesbertz}\ \emph {et~al.}(2008)\citenamefont
  {Giesbertz}, \citenamefont {Baerends},\ and\ \citenamefont
  {Gritsenko}}]{GiesbertzBaerendsGritsenko2008}%
  \BibitemOpen
  \bibfield  {author} {\bibinfo {author} {\bibfnamefont {K.~J.~H.}\
  \bibnamefont {Giesbertz}}, \bibinfo {author} {\bibfnamefont {E.~J.}\
  \bibnamefont {Baerends}}, \ and\ \bibinfo {author} {\bibfnamefont {O.~V.}\
  \bibnamefont {Gritsenko}},\ }\href {\doibase 10.1103/PhysRevLett.101.033004}
  {\bibfield  {journal} {\bibinfo  {journal} {Phys. Rev. Lett.}\ }\textbf
  {\bibinfo {volume} {101}},\ \bibinfo {pages} {033004} (\bibinfo {year}
  {2008})}\BibitemShut {NoStop}%
\bibitem [{\citenamefont {Giesbertz}\ \emph {et~al.}(2009)\citenamefont
  {Giesbertz}, \citenamefont {Pernal}, \citenamefont {Gritsenko},\ and\
  \citenamefont {Baerends}}]{GiesbertzPernalGritsenko2009}%
  \BibitemOpen
  \bibfield  {author} {\bibinfo {author} {\bibfnamefont {K.~J.~H.}\
  \bibnamefont {Giesbertz}}, \bibinfo {author} {\bibfnamefont {K.}~\bibnamefont
  {Pernal}}, \bibinfo {author} {\bibfnamefont {O.~V.}\ \bibnamefont
  {Gritsenko}}, \ and\ \bibinfo {author} {\bibfnamefont {E.~J.}\ \bibnamefont
  {Baerends}},\ }\href {\doibase 10.1063/1.3079821} {\bibfield  {journal}
  {\bibinfo  {journal} {J. Chem. Phys.}\ }\textbf {\bibinfo {volume} {130}},\
  \bibinfo {pages} {114104} (\bibinfo {year} {2009})}\BibitemShut {NoStop}%
\bibitem [{\citenamefont {Giesbertz}\ \emph {et~al.}(2010)\citenamefont
  {Giesbertz}, \citenamefont {Gritsenko},\ and\ \citenamefont
  {Baerends}}]{GiesbertzGritsenkoBaerends2010a}%
  \BibitemOpen
  \bibfield  {author} {\bibinfo {author} {\bibfnamefont {K.~J.~H.}\
  \bibnamefont {Giesbertz}}, \bibinfo {author} {\bibfnamefont {O.~V.}\
  \bibnamefont {Gritsenko}}, \ and\ \bibinfo {author} {\bibfnamefont {E.~J.}\
  \bibnamefont {Baerends}},\ }\href {\doibase 10.1103/PhysRevLett.105.013002}
  {\bibfield  {journal} {\bibinfo  {journal} {Phys. Rev. Lett.}\ }\textbf
  {\bibinfo {volume} {105}},\ \bibinfo {pages} {013002} (\bibinfo {year}
  {2010})}\BibitemShut {NoStop}%
\bibitem [{\citenamefont {Pernal}\ \emph {et~al.}(2007)\citenamefont {Pernal},
  \citenamefont {Gritsenko},\ and\ \citenamefont
  {Baerends}}]{PernalGritsenkoBaerends2007}%
  \BibitemOpen
  \bibfield  {author} {\bibinfo {author} {\bibfnamefont {K.}~\bibnamefont
  {Pernal}}, \bibinfo {author} {\bibfnamefont {O.}~\bibnamefont {Gritsenko}}, \
  and\ \bibinfo {author} {\bibfnamefont {E.~J.}\ \bibnamefont {Baerends}},\
  }\href {\doibase 10.1103/PhysRevA.75.012506} {\bibfield  {journal} {\bibinfo
  {journal} {Phys. Rev. A}\ }\textbf {\bibinfo {volume} {75}},\ \bibinfo
  {pages} {012506} (\bibinfo {year} {2007})}\BibitemShut {NoStop}%
\bibitem [{\citenamefont {van Leeuwen}(2003)}]{Leeuwen2003}%
  \BibitemOpen
  \bibfield  {author} {\bibinfo {author} {\bibfnamefont {R.}~\bibnamefont {van
  Leeuwen}},\ }in\ \href {\doibase 10.1016/S0065-3276(03)43002-5} {\emph
  {\bibinfo {booktitle} {Adv. Quant. Chem.}}},\ Vol.~\bibinfo {volume} {43},\
  \bibinfo {editor} {edited by\ \bibinfo {editor} {\bibfnamefont {J.~R.}\
  \bibnamefont {Sabin}}\ and\ \bibinfo {editor} {\bibfnamefont {E.~J.}\
  \bibnamefont {Braendas}}}\ (\bibinfo  {publisher} {Academic Press},\ \bibinfo
  {year} {2003})\ p.~\bibinfo {pages} {25}\BibitemShut {NoStop}%
\bibitem [{\citenamefont {Gilbert}(1975)}]{Gilbert1975}%
  \BibitemOpen
  \bibfield  {author} {\bibinfo {author} {\bibfnamefont {T.~L.}\ \bibnamefont
  {Gilbert}},\ }\href {\doibase 10.1103/PhysRevB.12.2111} {\bibfield  {journal}
  {\bibinfo  {journal} {Phys. Rev. B}\ }\textbf {\bibinfo {volume} {12}},\
  \bibinfo {pages} {2111} (\bibinfo {year} {1975})}\BibitemShut {NoStop}%
\bibitem [{\citenamefont {Fetter}\ and\ \citenamefont
  {Walecka}(2003)}]{FetterWalecka1971}%
  \BibitemOpen
  \bibfield  {author} {\bibinfo {author} {\bibfnamefont {A.~L.}\ \bibnamefont
  {Fetter}}\ and\ \bibinfo {author} {\bibfnamefont {J.~D.}\ \bibnamefont
  {Walecka}},\ }\href@noop {} {\emph {\bibinfo {title} {Quantum Theory of
  Many-Particle Systems}}}\ (\bibinfo  {publisher} {Dover Publiations, Inc.},\
  \bibinfo {year} {2003})\BibitemShut {NoStop}%
\bibitem [{\citenamefont {Lehmann}(1954)}]{Lehmann1954}%
  \BibitemOpen
  \bibfield  {author} {\bibinfo {author} {\bibfnamefont {H.}~\bibnamefont
  {Lehmann}},\ }\href {\doibase 10.1007/BF02783624} {\bibfield  {journal}
  {\bibinfo  {journal} {Nuovo Cimento}\ }\textbf {\bibinfo {volume} {11}},\
  \bibinfo {pages} {342} (\bibinfo {year} {1954})}\BibitemShut {NoStop}%
\bibitem [{Note1()}]{Note1}%
  \BibitemOpen
  \bibinfo {note} {The sum runs over a complete set of states, so also includes
  a possible continuum where the sum should be interpreted as an
  integral.}\BibitemShut {Stop}%
\bibitem [{Note2()}]{Note2}%
  \BibitemOpen
  \bibinfo {note} {In Ref.~\cite {RuggenthalerPenzLeeuwen2015} it is stated
  that the condition $\delta v_j(t) \in C^1$ can probably be weakened to
  Lipschitz continuity. This is still sufficient for our argument, since we
  only need continuity. A milder version of the Schr\IeC {\"o}dinger equation
  would allow for more general potentials in some $L^p$ spaces in time~\cite
  {PenzRuggenthaler2015, RuggenthalerPenzLeeuwen2015}. In that case, however,
  potentials which only differ at a set of zero measure would be considered
  equivalent.}\BibitemShut {Stop}%
\bibitem [{Note3()}]{Note3}%
  \BibitemOpen
  \bibinfo {note} {The matrix elements {$\delimiter 69640972 {\psi
  _i}\delimiter "526930B \delimiter "426830A {\psi _i}|{\protect \mathbf
  {r}}|{2s}\delimiter "526930B $} have been evaluated by calculating the
  corresponding integral.}\BibitemShut {Stop}%
\bibitem [{Note4()}]{Note4}%
  \BibitemOpen
  \bibinfo {note} {The momentum operator gives {$-\protect \mathrm {i}\partial
  _x\delimiter 69640972 {2s}\delimiter "526930B = 0$} in the limited Hilbert
  space $\protect \mathcal {H}$, so is simply the zero-operator. Using
  $(-\protect \mathrm {i}\partial _x)^3$ avoids such a pathological
  operator.}\BibitemShut {Stop}%
\bibitem [{\citenamefont {Hohenberg}\ and\ \citenamefont
  {Kohn}(1964)}]{HohenbergKohn1964}%
  \BibitemOpen
  \bibfield  {author} {\bibinfo {author} {\bibfnamefont {P.}~\bibnamefont
  {Hohenberg}}\ and\ \bibinfo {author} {\bibfnamefont {W.}~\bibnamefont
  {Kohn}},\ }\href {\doibase 10.1103/PhysRev.136.B864} {\bibfield  {journal}
  {\bibinfo  {journal} {Phys. Rev.}\ }\textbf {\bibinfo {volume} {136}},\
  \bibinfo {pages} {B864} (\bibinfo {year} {1964})}\BibitemShut {NoStop}%
\bibitem [{\citenamefont {Reed}\ and\ \citenamefont
  {Simon}(1980)}]{ReedSimon1980}%
  \BibitemOpen
  \bibfield  {author} {\bibinfo {author} {\bibfnamefont {M.}~\bibnamefont
  {Reed}}\ and\ \bibinfo {author} {\bibfnamefont {B.}~\bibnamefont {Simon}},\
  }\href@noop {} {\emph {\bibinfo {title} {Functional Analysis}}},\ \bibinfo
  {series} {Methods of Modern Mathematical Physics}\ No.~\bibinfo {number} {1}\
  (\bibinfo  {publisher} {Academic Press, Inc.},\ \bibinfo {address} {1250
  Sixth Avenue, San Diego, California 92101},\ \bibinfo {year}
  {1980})\BibitemShut {NoStop}%
\bibitem [{\citenamefont {Lieb}(1983)}]{Lieb1983}%
  \BibitemOpen
  \bibfield  {author} {\bibinfo {author} {\bibfnamefont {E.~H.}\ \bibnamefont
  {Lieb}},\ }\href {\doibase 10.1002/qua.560240302} {\bibfield  {journal}
  {\bibinfo  {journal} {Int. J. Quant. Chem.}\ }\textbf {\bibinfo {volume}
  {24}},\ \bibinfo {pages} {243} (\bibinfo {year} {1983})}\BibitemShut
  {NoStop}%
\bibitem [{\citenamefont {Lammert}(2015)}]{Lammert2015}%
  \BibitemOpen
  \bibfield  {author} {\bibinfo {author} {\bibfnamefont {P.~E.}\ \bibnamefont
  {Lammert}},\ }\href@noop {} {\enquote {\bibinfo {title} {Hohenberg--kohn
  redux},}\ } (\bibinfo {year} {2015}),\ \Eprint
  {http://arxiv.org/abs/1412.3876} {arXiv:1412.3876} \BibitemShut {NoStop}%
\bibitem [{Note5()}]{Note5}%
  \BibitemOpen
  \bibinfo {note} {An example is the hydrogen atom where the $1s$ state is
  excluded from the Hilbert space. Choose the $2p_x$ orbital as an initial
  state. The local potential which would produce only a component in the $2p_y$
  state would be $y/x$ which is infinite in the whole plane orthogonal to the
  $x$-axis.}\BibitemShut {Stop}%
\bibitem [{\citenamefont {Reed}\ and\ \citenamefont
  {Simon}(1975)}]{ReedSimon1975}%
  \BibitemOpen
  \bibfield  {author} {\bibinfo {author} {\bibfnamefont {M.}~\bibnamefont
  {Reed}}\ and\ \bibinfo {author} {\bibfnamefont {B.}~\bibnamefont {Simon}},\
  }\href@noop {} {\emph {\bibinfo {title} {Fourier Analysis,
  Self-Adjointness}}},\ \bibinfo {series} {Methods of Modern Mathematical
  Physics}, Vol.~\bibinfo {volume} {2}\ (\bibinfo  {publisher} {Academic
  Press},\ \bibinfo {year} {1975})\BibitemShut {NoStop}%
\bibitem [{\citenamefont {Kato}(1957)}]{Kato1957}%
  \BibitemOpen
  \bibfield  {author} {\bibinfo {author} {\bibfnamefont {T.}~\bibnamefont
  {Kato}},\ }\href {\doibase 10.1002/cpa.3160100201} {\bibfield  {journal}
  {\bibinfo  {journal} {Commun. Pure Appl. Math.}\ }\textbf {\bibinfo {volume}
  {10}},\ \bibinfo {pages} {151} (\bibinfo {year} {1957})}\BibitemShut
  {NoStop}%
\bibitem [{\citenamefont {von Barth}\ and\ \citenamefont
  {Hedin}(1972)}]{BarthHedin1972}%
  \BibitemOpen
  \bibfield  {author} {\bibinfo {author} {\bibfnamefont {U.}~\bibnamefont {von
  Barth}}\ and\ \bibinfo {author} {\bibfnamefont {L.}~\bibnamefont {Hedin}},\
  }\href {\doibase 10.1088/0022-3719/5/13/012} {\bibfield  {journal} {\bibinfo
  {journal} {J. Phys. C}\ }\textbf {\bibinfo {volume} {5}},\ \bibinfo {pages}
  {1629} (\bibinfo {year} {1972})}\BibitemShut {NoStop}%
\bibitem [{\citenamefont {Eschrig}\ and\ \citenamefont
  {Pickett}(2001)}]{EschrigPickett2001}%
  \BibitemOpen
  \bibfield  {author} {\bibinfo {author} {\bibfnamefont {H.}~\bibnamefont
  {Eschrig}}\ and\ \bibinfo {author} {\bibfnamefont {W.}~\bibnamefont
  {Pickett}},\ }\href {\doibase 10.1016/S0038-1098(01)00053-9} {\bibfield
  {journal} {\bibinfo  {journal} {Solid State Commun.}\ }\textbf {\bibinfo
  {volume} {118}},\ \bibinfo {pages} {123} (\bibinfo {year}
  {2001})}\BibitemShut {NoStop}%
\bibitem [{\citenamefont {L{\"o}wdin}(1955)}]{Lowdin1955}%
  \BibitemOpen
  \bibfield  {author} {\bibinfo {author} {\bibfnamefont {P.-O.}\ \bibnamefont
  {L{\"o}wdin}},\ }\href {\doibase 10.1103/PhysRev.97.1474} {\bibfield
  {journal} {\bibinfo  {journal} {Phys. Rev.}\ }\textbf {\bibinfo {volume}
  {97}},\ \bibinfo {pages} {1474} (\bibinfo {year} {1955})}\BibitemShut
  {NoStop}%
\bibitem [{\citenamefont {Friesecke}(2003)}]{Friesecke2003}%
  \BibitemOpen
  \bibfield  {author} {\bibinfo {author} {\bibfnamefont {G.}~\bibnamefont
  {Friesecke}},\ }\href {\doibase 10.1098/rspa.2002.1027} {\bibfield  {journal}
  {\bibinfo  {journal} {Proc. R. Soc. London A}\ }\textbf {\bibinfo {volume}
  {459}},\ \bibinfo {pages} {47} (\bibinfo {year} {2003})}\BibitemShut
  {NoStop}%
\bibitem [{\citenamefont {Giesbertz}\ and\ \citenamefont {van
  Leeuwen}(2013{\natexlab{a}})}]{GiesbertzLeeuwen2013a}%
  \BibitemOpen
  \bibfield  {author} {\bibinfo {author} {\bibfnamefont {K.~J.~H.}\
  \bibnamefont {Giesbertz}}\ and\ \bibinfo {author} {\bibfnamefont
  {R.}~\bibnamefont {van Leeuwen}},\ }\href {\doibase 10.1063/1.4820419}
  {\bibfield  {journal} {\bibinfo  {journal} {J. Chem. Phys.}\ }\textbf
  {\bibinfo {volume} {139}},\ \bibinfo {pages} {104109} (\bibinfo {year}
  {2013}{\natexlab{a}})}\BibitemShut {NoStop}%
\bibitem [{\citenamefont {Giesbertz}\ and\ \citenamefont {van
  Leeuwen}(2013{\natexlab{b}})}]{GiesbertzLeeuwen2013b}%
  \BibitemOpen
  \bibfield  {author} {\bibinfo {author} {\bibfnamefont {K.~J.~H.}\
  \bibnamefont {Giesbertz}}\ and\ \bibinfo {author} {\bibfnamefont
  {R.}~\bibnamefont {van Leeuwen}},\ }\href {\doibase 10.1063/1.4820418}
  {\bibfield  {journal} {\bibinfo  {journal} {J. Chem. Phys.}\ }\textbf
  {\bibinfo {volume} {139}},\ \bibinfo {pages} {104110} (\bibinfo {year}
  {2013}{\natexlab{b}})}\BibitemShut {NoStop}%
\bibitem [{\citenamefont {Giesbertz}\ and\ \citenamefont {van
  Leeuwen}(2014)}]{GiesbertzLeeuwen2014}%
  \BibitemOpen
  \bibfield  {author} {\bibinfo {author} {\bibfnamefont {K.~J.~H.}\
  \bibnamefont {Giesbertz}}\ and\ \bibinfo {author} {\bibfnamefont
  {R.}~\bibnamefont {van Leeuwen}},\ }\href {\doibase 10.1063/1.4875338}
  {\bibfield  {journal} {\bibinfo  {journal} {J. Chem. Phys.}\ }\textbf
  {\bibinfo {volume} {140}},\ \bibinfo {pages} {184108} (\bibinfo {year}
  {2014})}\BibitemShut {NoStop}%
\bibitem [{\citenamefont {L{\"o}wdin}\ and\ \citenamefont
  {Shull}(1956)}]{LowdinShull1956}%
  \BibitemOpen
  \bibfield  {author} {\bibinfo {author} {\bibfnamefont {P.-O.}\ \bibnamefont
  {L{\"o}wdin}}\ and\ \bibinfo {author} {\bibfnamefont {H.}~\bibnamefont
  {Shull}},\ }\href {\doibase 10.1103/PhysRev.101.1730} {\bibfield  {journal}
  {\bibinfo  {journal} {Phys. Rev.}\ }\textbf {\bibinfo {volume} {101}},\
  \bibinfo {pages} {1730} (\bibinfo {year} {1956})}\BibitemShut {NoStop}%
\bibitem [{\citenamefont {Cioslowski}\ \emph {et~al.}(2002)\citenamefont
  {Cioslowski}, \citenamefont {Pernal},\ and\ \citenamefont
  {Ziesche}}]{CioslowskiPernalZiesche2002}%
  \BibitemOpen
  \bibfield  {author} {\bibinfo {author} {\bibfnamefont {J.}~\bibnamefont
  {Cioslowski}}, \bibinfo {author} {\bibfnamefont {K.}~\bibnamefont {Pernal}},
  \ and\ \bibinfo {author} {\bibfnamefont {P.}~\bibnamefont {Ziesche}},\ }\href
  {\doibase 10.1063/1.1516804} {\bibfield  {journal} {\bibinfo  {journal} {J.
  Chem. Phys.}\ }\textbf {\bibinfo {volume} {117}},\ \bibinfo {pages} {9560}
  (\bibinfo {year} {2002})}\BibitemShut {NoStop}%
\bibitem [{\citenamefont {Giesbertz}(2010)}]{PhD-Giesbertz2010}%
  \BibitemOpen
  \bibfield  {author} {\bibinfo {author} {\bibfnamefont {K.~J.~H.}\
  \bibnamefont {Giesbertz}},\ }\emph {\bibinfo {title} {Time-Dependent One-Body
  Reduced Density Matrix Functional Theory; Adiabatic Approximations and
  Beyond}},\ \href@noop {} {Ph.D. thesis},\ \bibinfo  {school} {{V}rije
  {U}niversiteit}, \bibinfo {address} {De Boelelaan 1105, Amsterdam, The
  Netherlands} (\bibinfo {year} {2010})\BibitemShut {NoStop}%
\bibitem [{\citenamefont {Rapp}\ \emph {et~al.}(2014)\citenamefont {Rapp},
  \citenamefont {Brics},\ and\ \citenamefont {Bauer}}]{RappBricsBauer2014}%
  \BibitemOpen
  \bibfield  {author} {\bibinfo {author} {\bibfnamefont {J.}~\bibnamefont
  {Rapp}}, \bibinfo {author} {\bibfnamefont {M.}~\bibnamefont {Brics}}, \ and\
  \bibinfo {author} {\bibfnamefont {D.}~\bibnamefont {Bauer}},\ }\href
  {\doibase 10.1103/PhysRevA.90.012518} {\bibfield  {journal} {\bibinfo
  {journal} {Phys. Rev. A}\ }\textbf {\bibinfo {volume} {90}},\ \bibinfo
  {pages} {012518} (\bibinfo {year} {2014})}\BibitemShut {NoStop}%
\bibitem [{Note6()}]{Note6}%
  \BibitemOpen
  \bibinfo {note} {If $\Psi _0$ is a singlet state,~\protect \textup {\hbox
  {\mathsurround \z@ \protect \normalfont (\ignorespaces \ref
  {eq:pairedZeroResponse}\unskip \@@italiccorr )}} are triplet
  operators.}\BibitemShut {Stop}%
\bibitem [{\citenamefont {Penz}\ and\ \citenamefont
  {Ruggenthaler}(2015)}]{PenzRuggenthaler2015}%
  \BibitemOpen
  \bibfield  {author} {\bibinfo {author} {\bibfnamefont {M.}~\bibnamefont
  {Penz}}\ and\ \bibinfo {author} {\bibfnamefont {M.}~\bibnamefont
  {Ruggenthaler}},\ }\href {\doibase 10.1063/1.4916390} {\bibfield  {journal}
  {\bibinfo  {journal} {J. Chem. Phys.}\ }\textbf {\bibinfo {volume} {142}},\
  \bibinfo {pages} {124113} (\bibinfo {year} {2015})}\BibitemShut {NoStop}%
\end{thebibliography}%

\end{document}